\documentclass[aps,prl,twocolumn]{revtex4}

\usepackage{graphicx}
\usepackage{bm}        
\usepackage{amsmath}
\usepackage{amssymb}   
\usepackage{array}
\usepackage[caption=false]{subfig}
\usepackage[bookmarks=false,linkcolor=blue,urlcolor=blue,colorlinks,citecolor=blue]{hyperref}
\makeatletter

\begin{document}

\title{A Gapless Symmetry-Protected Topological Phase of Fermions in One Dimension}

\author{Anna Keselman and Erez Berg}
\affiliation{Department of Condensed Matter Physics, Weizmann Institute of Science, Rehovot, Israel 76100}

\date{\today}

\begin{abstract}
We consider a one-dimensional, time-reversal-invariant system with
attractive interactions and spin-orbit coupling. Such a system is gapless due to
the strong quantum fluctuations of the superconducting order parameter.
However, we show that a sharply defined topological phase with protected,
exponentially localized edge states exists. If one of the spin
components is conserved, the protection of the edge modes can be understood
as a consequence of the presence of a spin gap. 
In the more general case, the localization of the edge states arises from
a gap to single particle excitations in the bulk. We consider specific
microscopic models and demonstrate both analytically
and numerically (using density matrix renormalization group calculations)
that they can support the topologically non-trivial phase.
\end{abstract}
\maketitle

\section{Introduction}

Topological phases of matter are typically characterized by a gapped
bulk spectrum, and protected gapless edge states with unique properties.
The existence of a finite energy gap in the bulk plays a crucial role
in the topological protection of the edge states. This therefore raises
the question whether a topological phase, i.e. a phase with protected
exponentially localized edges states, can exist in a gapless system~\cite{Bonderson2013}.

Superconductors have been shown to host a variety of topological phases~\cite{ReadGreen2000, kitaev2009periodic, Ryu2010, volovik2009universe, HasanKane2010, QiZhang2011}, depending on the symmetries and the dimensionality of the system. In one spatial dimension, such
a phase hosts protected edge modes, termed Majorana bound states,
which are in particular interesting due to their non-Abelian exchange
statistics. Realization of this phase requires
proximity coupling a one-dimensional system to a bulk three-dimensional superconductor~\cite{Kitaev2001,Oreg2010,Lutchyn2010,Alicea2012,Beenakker2013}. In truly one-dimensional systems
with intrinsic attractive interactions, strong quantum fluctuations
of the superconducting order parameter leave the system gapless. E.g., this situation can arise if a quantum wire is coupled to a superconducting wire. It
was shown that topological protection is then much weaker in general~\cite{Fidkowski2011,Sau2011} (the edge states are generically only power-law
localized, rather than exponentially localized) and depends on the microscopics of the
system~\cite{ruhman2014topological}.

In the presence of time-reversal symmetry a different kind of topological
superconductors can be realized~\cite{Qi2009, Fu2010, Deng2012, Nakosai2012, Seradjeh2012, Wong2012, Zhang2013, Nakosai2013, Keselman2013, Haim2014, Gaidamauskas2014}. If the bulk is fully
gapped, these phases host a Kramers pair of Majorana bound
states~\cite{Keselman2013,Liu2014,WolmsStern2014, Zhao2014}.

In this work, we show that one-dimensional time reversal symmetric systems can support a well-defined topological superconducting phase, even when the particle number is conserved. The two necessary ingredients to realize this phase are attractive effective interactions and spin-orbit coupling. We start by giving a field theoretical
argument for the existence of the topological phase, distinct from a
trivial one. We show that the boundary between the topological and
the trivial phases hosts exponentially localized edge states and discuss
their properties. The edges exhibit an anomalous relation between time reversal and the local fermion parity operator, just like in the fully gapped case. We then demonstrate the existence of the topological
phase and its properties numerically. To this end we consider a simple model
that exhibits a phase transition between the trivial and the topological
phases and study it using the density matrix renormalization group
(DMRG) technique. Finally, we consider a system consisting of a semiconducting
wire with spin-orbit coupling and repulsive interactions coupled to
a superconducting wire, and show using bosonization and weak coupling
renormalization group (RG) that it can be driven into the gapless topological phase.

\section{Field theory of gapless time-reversal-invariant topological superconductors}

\subsection{Spin conserving case}

We begin by considering an interacting, time reversal invariant one-dimensional electron gas (1DEG), described at low energies as a Luttinger liquid.
In bosonized language, the Hamiltonian can be written as
\begin{equation}
H_0 = \underset{\alpha=\rho,\sigma}{\sum} \frac{u_{\alpha}}{2\pi} \int \left( K_{\alpha}\left(\partial_x\theta_{\alpha} \right)^2 + \frac{1}{K_{\alpha}}\left(\partial_x\phi_{\alpha}\right)^2 \right) dx,
\end{equation}
where $\rho,\sigma$ correspond to the charge and spin degrees of
freedom, $\rho(x) = -\frac{1}{\pi}\partial_{x}\phi_{\rho}$ and $s^z(x) = -\frac{1}{2\pi} \partial_{x}\phi_{\sigma}$ are the
charge and spin densities, respectively, and $\frac{1}{\pi}\theta_{\alpha}$ is
the field conjugate to $\phi_{\alpha}$,
$\left[\phi_{\alpha}\left(x\right), \theta_{\alpha}\left(x'\right)\right]=i \pi \Theta \left(x'-x\right)$, where $\Theta(x)$ is the Heaviside step function.

The relation to fermionic operators, describing modes linearized around the Fermi momentum $k_F$, is given by
\begin{equation}
R\left(L\right)_s=\frac{U_s}{\sqrt{2\pi a}}e^{-i\left(\pm\frac{1}{2}\phi_{\rho}-\theta_{\rho}+s\left(\pm\frac{1}{2}\phi_{\sigma}-\theta_{\sigma}\right)\right)}.
\end{equation}
where $s$ denotes the spin of the fermion, $U_{s}$ are the Klein factors that impose the anti-commutation relations between the different spin species, $+$ ($-$) signs correspond to R(L), representing right(left) movers, respectively, and $a$ is the short distance cutoff in the theory~\cite{GiamarchiBook}.
Back-scattering processes give rise to cosine terms that can gap out some of the modes in the system.
We assume the system to be at a generic filling, such that there are no relevant umklapp processes.

Our system is symmetric under time-reversal (TR) symmetry, denoted by $\mathcal{T}$ (class DIII in the Zirnbauer-Altland classification~\cite{AltlandZirnbauer1997}, such that $\mathcal{T}^2=-1$). Under time reversal, $R_{\uparrow} \rightarrow L_{\downarrow}$, $L_{\downarrow}\rightarrow -R_{\uparrow}$, $L_{\uparrow} \rightarrow R_{\downarrow}$, and $R_{\downarrow} \rightarrow -L_{\uparrow}$. These relations correspond to the following transformations of the bosonic fields: $\theta_\rho \rightarrow -\theta_\rho$, $\phi_\rho \rightarrow \phi_\rho$, $\theta_\sigma \rightarrow \theta_\sigma$, $\phi_\sigma \rightarrow -\phi_\sigma$, $U_\uparrow \rightarrow U_\downarrow$, and $U_\downarrow \rightarrow -U_\uparrow$. These transformation rules reproduce the correct behavior of the spin and charge densities and currents under TR.

We consider a system with spin-orbit coupling, and hence no SU(2) spin symmetry. Let us assume, for simplicity, that the spin is conserved along one direction (e.g. the $z$ direction, such that $S_z$ is conserved). This condition will be relaxed later on. The most general four-fermion back-scattering interaction
consistent with time-reversal symmetry is 
\begin{equation}
g\left(R_{\uparrow}^{\dagger}L_{\uparrow}^{\phantom{\dagger}}L_{\downarrow}^{\dagger}R_{\downarrow}^{\phantom{\dagger}}+h.c.\right) = \frac{g}{2 \pi^2 a^2}\cos\left(2\phi_{\sigma}\right),
\end{equation}
where the coupling $g$ is a real number. Note that such a cosine term is invariant under $\mathcal{T}$, according to the transformation rule of  $\phi_{\sigma}$ above. (Higher order processes
are also possible, but are less relevant in the RG sense.)
If this cosine term is relevant, it opens a spin gap,
driving the system into a ``Luther-Emery phase''~\cite{LutherEmery1974} with only one gapless (charge) mode. The sign of $g$ determines the nature of the resulting
phase. To understand this recall that the spin-singlet and spin-triplet
pairing order parameters are given by
\begin{equation}
\begin{split}
O_{\rm SS} =R_{\uparrow}^{\dagger}L_{\downarrow}^{\dagger}+L_{\uparrow}^{\dagger}R_{\downarrow}^{\dagger}\propto U_{\uparrow}U_{\downarrow} e^{-2i\theta_{\rho}}\cos\left(\phi_{\sigma}\right), \\
O^\mathrm{z}_{\rm TS} =R_{\uparrow}^{\dagger}L_{\downarrow}^{\dagger}-L_{\uparrow}^{\dagger}R_{\downarrow}^{\dagger}\propto U_{\uparrow}U_{\downarrow} e^{-2i\theta_{\rho}}\sin\left(\phi_{\sigma}\right),
\end{split}
\end{equation}
while the spin-density wave and charge-density wave order parameters are given by
\begin{equation}
\begin{split}
O_{\rm CDW} =R_{\uparrow}^{\dagger}L_{\uparrow}^{\phantom{\dagger}}+L_{\downarrow}^{\dagger}R_{\downarrow}^{\phantom{\dagger}}\propto e^{-i\phi_{\rho}}{\cos}\left(\phi_{\sigma}\right), \\
O^\mathrm{z}_{\rm SDW} =R_{\uparrow}^{\dagger}L_{\uparrow}^{\phantom{\dagger}} - L_{\downarrow}^{\dagger}R_{\downarrow}^{\phantom{\dagger}}\propto e^{-i\phi_{\rho}}{\sin}\left(\phi_{\sigma}\right). \\
\end{split}
\label{eq:OrderParamDW}
\end{equation}
For $g<0$, the cosine term pins the field $\phi_{\sigma}$ to zero (or equivalently
to any integer multiple of $\pi$, i.e. $\pi n$
with $n\in\mathbb{Z}$), resulting in the dominant superconducting
correlations being of the spin-singlet order parameter. (As the charge
sector remains gapless, these pairing correlations decay with a power
law dictated by the Luttinger parameter $K_{\rho}$ and no true long
range order can develop). For $g>0$, $\phi_{\sigma}$ is pinned to
$\pi\left(n+\frac{1}{2}\right)$, where $n\in\mathbb{Z}$,
and the dominant superconducting correlations are now of the spin-triplet
order parameter. These two cases correspond to two distinct phases; in
order to go between them without breaking time-reversal symmetry,
the coupling $g$ has to cross zero, resulting in the closing of the spin
gap~\cite{comment-sg}. We refer to these phases as \emph{trivial} and \emph{topological}
respectively, in analogy with the fully gapped case. When the superconducting
order parameter is conventional s-wave, we expect the system to be
in the trivial phase, while for a p-wave order parameter the system
is topological. This identification is in agreement with the vacuum
being in the trivial phase, as a large back-scattering potential at
the end of the system pins $\phi_{\sigma}$ to zero.

The phase diagram of an interacting 1DEG with spin-orbit coupling, and the possibility of a phase with dominant triplet superconducting correlations was discussed in Refs.~\cite{GiamarchiSchulz1986,GiamarchiSchulz1988}, where it was demonstrated how this phase can arise in a model that describes certain quasi-one-dimensional organic conductors. However, to the best of our knowledge, the topological nature of this phase (manifested in its protected edge modes, as we argue below) has not been discussed.

Consider an edge of a topological system, or equivalently a boundary
between a trivial and a topological phase. Since the field $\phi_\sigma$ is pinned to $\pi n_1$ on one side of the boundary and to $\pi(n_2 + \frac{1}{2})$ on the other, where $n_{1,2}$ are integers, 
there must be a kink of minimal magnitude $\pm\frac{\pi}{2}$
in $\phi_{\sigma}$ across the boundary. Such a kink in $\phi_{\sigma}$
corresponds to an accumulation of spin:
\begin{equation}
S_z=\int s^z(x) dx=-\int\frac{1}{2\pi}\partial_x\phi_{\sigma}dx=\pm\frac{1}{4},
\end{equation}
i.e. half the spin of an electron. The same ``fractional spin'' appears at the edge of a time-reversal invariant
fully gapped topological superconductor~\cite{Keselman2013}.

\begin{figure}[ht]
\includegraphics[width=0.45\textwidth]{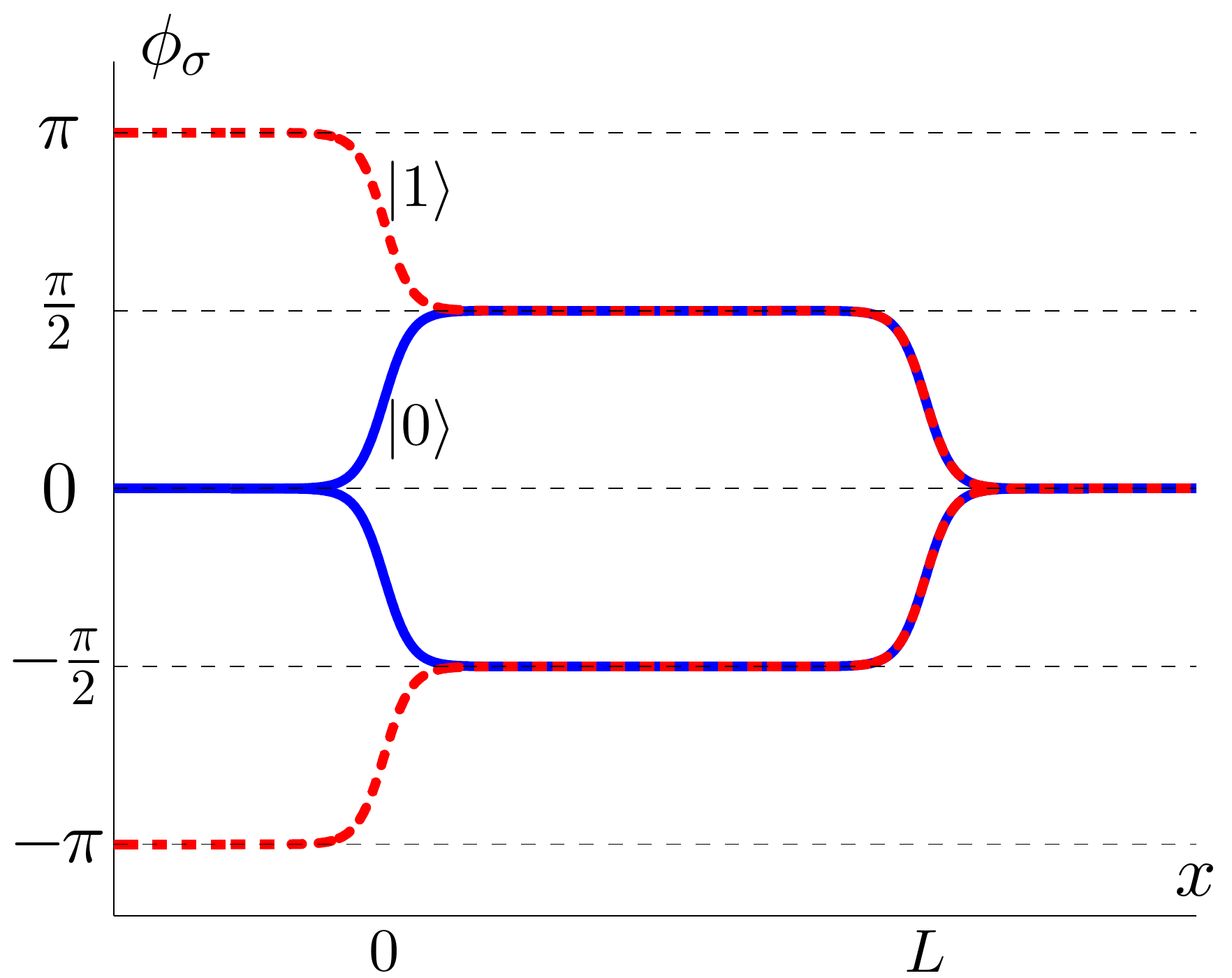}
\caption{Low energy configurations of $\phi_{\sigma}$ in a finite size gapless topological
superconductor. Configurations plotted in red dashed lines correspond
to total spin of $\pm\frac{1}{2}$ in the system, i.e. odd number
of particles, while configurations plotted in blue solid lines correspond
to zero total spin, i.e. even number of particles. To calculate the
local tunneling density of states we calculate the matrix element
of a single particle creation operator $\Psi_{\uparrow}^{\dagger}\left(x\right)$
between the states denoted by $\left|0\right\rangle $ and $\left|1\right\rangle $.}
\label{fig:4foldDegen}
\end{figure}

The lowest energy configurations of $\phi_{\sigma}$ for a topological
system of finite size are shown in Fig. \ref{fig:4foldDegen}. The two configurations
plotted in blue solid lines correspond to an even number of particles
in the system with total $\left\langle S_{z}\right\rangle =0$. These
configurations are degenerate up to a splitting exponential in
system size, as in the case of a fully gapped topological superconductor
with time-reversal symmetry. Adding a single particle in the bulk
of the system requires creating a kink of magnitude $\pm\pi$ (as the extra particles carries
spin $S_{z}=\pm\frac{1}{2}$).
Such a kink costs a finite amount of energy.
Adding a particle near the edge, however, only requires
flipping the direction of the $\frac{\pi}{2}$ kink at that edge without an extra energy
cost.
Since the number of particles in the system
is a good quantum number, adding a particle costs charging energy,
but this contribution decreases with the system size $L$ as $\frac{1}{L}$.
The resulting configurations with an odd number of particles and total
$\left\langle S_{z}\right\rangle =\pm\frac{1}{2}$ are given by 
red dashed lines in Fig. \ref{fig:4foldDegen}. (Note that the two
states with an odd number of particles are exactly degenerate,
in accord with Kramers' theorem.) We therefore expect that the
gapless topological phase will host single particle edge excitations which
cost zero energy in the thermodynamic limit. This is in sharp contrast
to the gapless trivial phase where no such edge excitations exist.

\begin{figure}[t]
\subfloat[]{\includegraphics[width=0.24\textwidth]{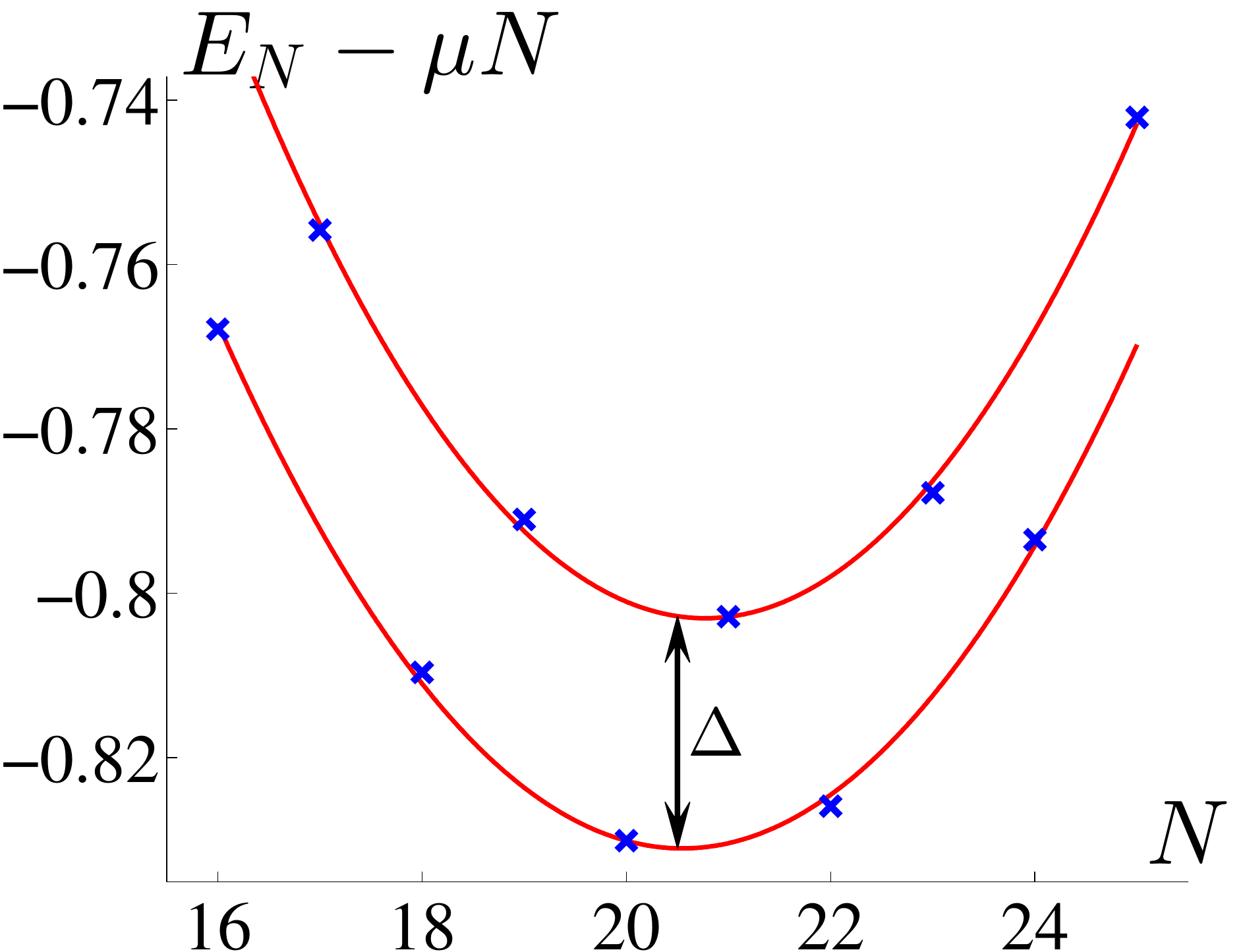}}
\subfloat[]{\includegraphics[width=0.24\textwidth]{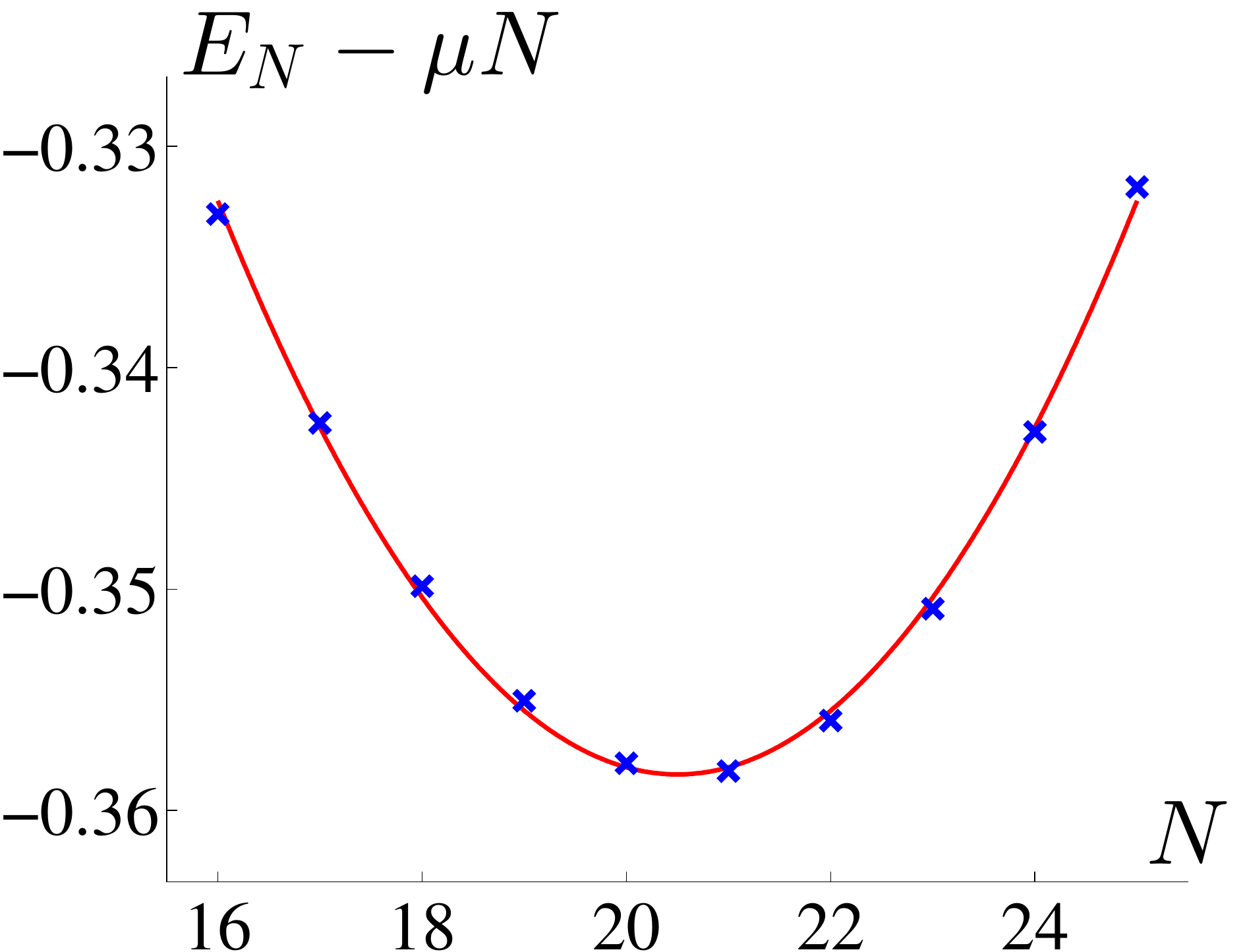}}
\caption{
The ground state energy of a system with open boundary conditions as the number of particles is varied for (a) a system in the trivial phase (b) a system in the topological phase. In both
cases the contribution of the chemical potential is subtracted, $\Delta$
indicates the single particle gap. In the topological phase a low energy state appears for an odd number of particles due to the presence of the low energy edge state. Note that a single particle gap is still present in the bulk of the system.
The data is obtained using DMRG study of the model given by Eq. \ref{eq:HToyModel} for a system of length $L=100$ sites with model parameters $t=1,\ U=-1$ and spin-orbit coupling $v=1$ in (a) and $v=0$ in (b). The chemical potential that is subtracted in both cases is found from a linear fit of $E_N$ to $N$. The red solid lines are fits to parabolas. The finite curvature of the parabolas is due to the charging energy, and depends on the size of the system.
}
\label{fig:SpectrumVsN}
\end{figure}

This distinction can be seen in Fig. \ref{fig:SpectrumVsN}, depicting the ground state energy of a system with open boundary conditions as a function of the number of particles in the two cases. For a system in the trivial phase, states with an odd number of particles lie on a parabola separated by $\Delta$, the single particle gap in the system, from the states with an even number of particles. For a system in the topological phase, an extra particle can be added at the edge of the system at a cost of the charging energy only.
The two states are distinguished by the pair binding energy in the system 
\begin{equation}
E_{B}=E_{2N+1}-\frac{1}{2}\left(E_{2N+2}+E_{2N}\right)
\label{eq:EBinding}
\end{equation}
as the size of the system is varied. Extrapolating to the infinite system size limit, we expect to obtain the single particle gap, ${\rm lim}_{L\rightarrow\infty}E_{B}=\Delta>0$, if the system is in the trivial phase. If the system is in the topological phase and with open boundary conditions, we expect ${\rm lim}_{L\rightarrow\infty}E_{B}$ to be zero. In contrast, in a system with periodic boundary conditions, both phases are characterized by a finite $E_{B}$ in the thermodynamic limit.

Experimentally, the existence of a low energy single particle edge state is reflected in the local tunneling density of states (TDOS)
\begin{equation}
\begin{split}
\frac{1}{\pi} & {\rm Im}G^{\rm R}\left(\omega,x\right) = \\
& \underset{n,s}{\sum}\delta\left(\omega-\omega_{n}\right)\left(\left|\left\langle 0\right|\Psi_{s}^{\dagger}\left(x\right)\left|n\right\rangle \right|^{2}+\left|\left\langle 0\right|\Psi_{s}\left(x\right)\left|n\right\rangle \right|^{2}\right),
\end{split}
\label{eq:TDOS}
\end{equation}
in the $\omega\rightarrow0$ limit. Here $\left|0\right>$ denotes the ground state of the system and $\left|n\right>$ an excited state at energy $\hbar\omega_n$. Assuming the number of particles in the ground state is even, the configuration of $\phi_{\sigma}$ in this state is given by one of the blue solid curves in Fig.
\ref{fig:4foldDegen}, e.g. the one denoted by $\left|0\right\rangle $.
The lowest energy excited state that contributes to the sum in
Eq. \ref{eq:TDOS} corresponds to a configuration of $\phi_{\sigma}$
denoted by $\left|1\right\rangle $.
Using the bosonized representation of $\Psi_{s}^{\dagger}\left(x\right)$
and a mode expansion for the bosonic fields, $\phi_{\rho,\sigma}$ and
$\theta_{\rho,\sigma}$, we obtain for the matrix element:
\begin{equation}
\left|\left\langle 1\right|\Psi_{\uparrow}^{\dagger}\left(x\right)\left|0\right\rangle \right|^{2}\sim \left(\frac{a}{L}\right)^{\frac{1}{K_{\rho}}}\left(\frac{a}{x}\right)^{\frac{\alpha}{2}}e^{-\frac{\pi}{2K_{\sigma}}\frac{x}{\xi}},
\label{eq:MatrixElement}
\end{equation}
where $\alpha=\frac{1}{K_{\rho}}+\frac{1}{4}K_{\rho}$, $\Delta_{\sigma}$ is the spin gap, and
$\xi=\frac{u_{\sigma}}{\Delta_{\sigma}}$ is the correlation length in the spin sector.
For details of the calculation see Appendix~\ref{sec:Appendix-TDOS}.

\subsection{Non-spin conserving case}

We now generalize the analysis of the previous section to the more generic case, in which none of the spin components are conserved.

\begin{figure}
\includegraphics[width=0.3\textwidth]{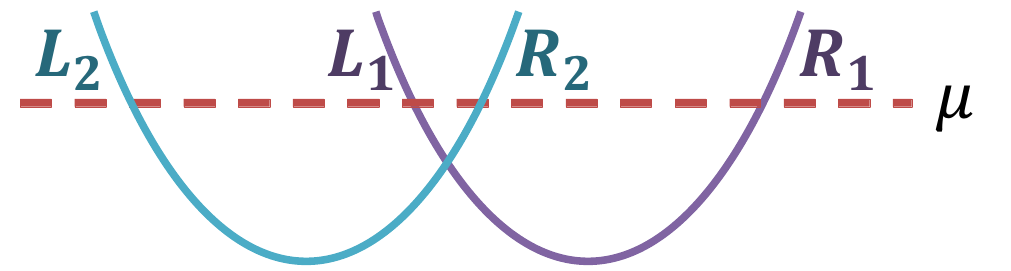}
\caption{Generic spectrum for a system without inversion symmetry. $R(L)_{1,2}$
denote the right(left) movers in each band. Under time-reversal symmetry
$R_{1}\rightarrow L_{2}$ and $R_{2}\rightarrow-L_{1}$.}
\label{fig:SpectrumNoInversion}
\end{figure}

Let us first consider a case where the system has no inversion center. Then, the single particle spectrum 
is generically of the form shown in Fig. \ref{fig:SpectrumNoInversion}. We linearize
the modes close to Fermi energy and denote by $R(L)_{1,2}$ the right
(left) movers in each band. Under time-reversal symmetry $R_{1}\rightarrow L_{2}$
and $R_{2}\rightarrow-L_{1}$. We can therefore assign a pseudo-spin
$\uparrow$, $\downarrow$ to the two bands, respectively, and write
the linearized fermionic modes in terms of the bosonic fields $\phi_{\rho,\sigma}$
and $\theta_{\rho,\sigma}$ exactly as before, where $\phi_{\sigma}$ is now related
to the density of the pseudo-spin. The only four-particle back-scattering
term allowed by time-reversal symmetry is $R_{1}^{\dagger}L_{1}^{\phantom{\dagger}}L_{2}^{\dagger}R_{2}^{\phantom{\dagger}}+h.c.$, or $\cos\left(2\phi_{\sigma}\right)$ in terms of the bosonic fields, and
all the analysis presented for the $S_{z}$ conserving case still holds.

If the system has an inversion center, the resulting phase diagram is richer. The case of spin anisotropic interactions in an inversion symmetric
system with time-reversal symmetry was analyzed in \cite{GiamarchiSchulz1988}. Here, we outline this analysis briefly, and comment on the topological nature of the various phases and their boundary properties.

After an appropriate rotation in spin space, the most general four-fermion inversion symmetric interaction can be written as
\begin{equation}
H_{\rm int}=g_s\Delta_s^{\dagger}\Delta_s^{\phantom{\dagger}}+\underset{\hat{d}=\hat{x},\hat{y},\hat{z}}{\sum}g^{\phantom{\dagger}}_{\hat{d}}\Delta_{\hat{d}}^{\dagger}\Delta_{\hat{d}}^{\phantom{\dagger}},
\end{equation}
where $\Delta_s = L^T\left( i s^y \right)R^{\phantom{T}}$ is a singlet order parameter, $\Delta_{\hat{d}} = L^T\left( i s^y \hat{d}\cdot \vec{s} \right)R^{\phantom{T}}$ is a time-reversal symmetric triplet order parameter characterized by the orientation of a headless $\hat{d}$ vector, and we have dropped terms that do not couple the right and left movers. Here $L(R)$ are spinors in spin space, $L(R)^T=\left(L(R)_{\uparrow}, L(R)_{\downarrow} \right)$. Note that inversion symmetry prohibits cross terms that couple the singlet and triplet order parameters.

The back-scattering part of this interaction can be written in terms of the bosonic fields as
\begin{equation}
H_{\rm int} = \frac{1}{2\pi^2a^2}\left\{ \left(g_{\hat{z}}-g_s\right)\cos\left(2\phi_{\sigma}\right) +\left(g_{\hat{y}}-g_{\hat{x}}\right)\cos\left(4\theta_{\sigma}\right) \right\}.
\end{equation}
The resulting phase diagram hosts four phases with dominant superconducting correlations. These correspond to pinning of either $\phi_{\sigma}$ or $\theta_{\sigma}$ with two possible values for each, depending on the sign of $\left(g_{\hat{z}}-g_s\right)$ or $\left(g_{\hat{y}}-g_{\hat{x}}\right)$, respectively. Let us denote the phase in which $\phi_\sigma=0$ by SS (dominant singlet superconducting correlations), and the phase with $\phi_\sigma=\pm\frac{\pi}{2}$ by TS$_\mathrm{z}$ (dominant triplet correlations with $\hat{d}\parallel \hat{z}$). The other two phases with $\theta_{\sigma}=n\frac{\pi}{2},\left(n+\frac{1}{2}\right)\frac{\pi}{2}$ 
will be denoted as TS$_\mathrm{x}$ and TS$_\mathrm{y}$, respectively.

The SS and TS$_\mathrm{z}$ phases are exactly the ones discussed in the context of an $S_z$ conserving system. These phases are distinct from each other (in the sense that they cannot be adiabatically connected without a phase transition in which the gap in the spin sector closes) 
as long as time reversal symmetry is preserved. Below, we argue that the TS$_\mathrm{x,y,z}$ can be adiabatically connected to each other without breaking time reversal symmetry. However, they are distinct as long as certain mirror symmetries, $M_{x,y,z}$ (where $M_x$ is a mirror symmetry that takes $x\rightarrow -x$, etc.) are maintained; in the presence of such a mirror symmetry, an interface between a pair of TS phases with different spin quantization axes hosts protected edge states.

To see that the TS$_\mathrm{x}$ and TS$_\mathrm{y}$ can be adiabatically connected without TR symmetry breaking, consider the following family of interaction Hamiltonians: $H'_{\mathrm{int}}(\alpha) = -\left|g'\right| \Delta^\dagger_{\hat{d}} \Delta_{\hat{d}}^{\phantom{\dagger}}$, where $\hat{d} = \cos(\alpha) \hat{x} + \sin(\alpha) \hat{y}$. In bosonized form, $H'_{\mathrm{int}}(\alpha) \propto -\left|g'\right| \cos\left(4(\theta_\sigma - \alpha/2) \right)$; therefore, as $\alpha$ varies from $0$ to $\frac{\pi}{2}$, the value to which $\theta_\sigma$ is pinned changes from $0$ to $\frac{\pi}{4}$, the values that correspond to the  TS$_\mathrm{x}$ and TS$_\mathrm{y}$ phases, respectively. Interpolating from the TS$_\mathrm{x}$ or TS$_\mathrm{y}$ phases to the TS$_\mathrm{z}$ phase is more difficult in the context of Abelian bosonization; nevertheless, by a rotation in spin space we can show that these phases can also be adiabatically connected in a similar fashion.

Note, however, that this path in Hamiltonian space defined above includes terms like $\Delta^\dagger_{\hat{x}} \Delta_{\hat{y}}^{\phantom{\dagger}}$, that break $M_x$ and $M_y$. In the presence of either of these mirror symmetries, such terms are forbidden, and the TS$_\mathrm{x}$ and TS$_\mathrm{y}$ phases are distinct from each other.

In order to understand the topological nature of the TS$_\mathrm{x}$ and TS$_\mathrm{y}$ phases, we examine the low energy spectrum in a system with open boundary conditions.
Let us consider a system of length $L$, with boundary conditions imposed by a strong back-scattering potential at both ends. The back-scattering term has the form
\begin{equation}
-\left|V_b\right| \underset{s=\uparrow,\downarrow}{\sum}R_s^{\dagger}L_s^{\phantom{\dagger}}+h.c.= -\frac{2 \left|V_b\right|}{\pi a}\cos\left(\phi_{\rho}\right)\cos\left(\phi_{\sigma}\right).
\label{eq:BoundaryConditions}
\end{equation}
Such a potential pins both $\phi_{\rho}$ and $\phi_{\sigma}$ to either an even or an odd multiple of $\pi$ at both ends of the system, $x=0$ and $L$. A system described by a state with $\theta_{\sigma}$ pinned to a specific value in the bulk can not satisfy these boundary conditions, due to the non-trivial commutation relations $\left[\phi_{\sigma}\left(0\right), \theta_{\sigma}\left(x\right)\right]=i\pi$, where $0<x<L$.

Moreover, the fermion parity 
$\mathcal{P}=e^{i \pi N}$, where $N=-\frac{1}{\pi} \left( \phi_{\rho}\left(L\right)-\phi_{\rho}\left(0\right) \right)$ is the total number of particles, is not well defined in such a state. 
To see this note that using the boundary conditions dictated by Eq.~\ref{eq:BoundaryConditions}, we can rewrite $\mathcal{P}$ in terms of the spin degrees of freedom as
\begin{equation}
\mathcal{P}=e^{-i \left(\phi_{\sigma}\left(L\right)-\phi_{\sigma}\left(0\right)\right)},
\label{eq:FermionParity}
\end{equation}
since 
$\left( \phi_{\rho}\left(L\right)-\phi_{\rho}\left(0\right) \right)=\left( \phi_{\sigma}\left(L\right)-\phi_{\sigma}\left(0\right) \right) \mod 2\pi$.
Therefore, the following anti-commutation relations hold $\mathcal{P}e^{i\theta_{\sigma}\left(x\right)}=-e^{i\theta_{\sigma}\left(x\right)} \mathcal{P}$.

Ground states of the, e.g. TS$_{\mathrm{x}}$ phase, with well defined fermion parity can be constructed in the following way: $\vert \psi_{0,\pm}  \rangle =  \left|\theta_{\sigma}=0\right> \pm \left|\theta_{\sigma}=\pi \right>$ and $\vert \psi_{\frac{\pi}{2} ,\pm}  \rangle =  \left|\theta_{\sigma}=\frac{\pi}{2}\right> \pm \left|\theta_{\sigma}=\frac{3\pi}{2}\right>$, where the symmetric (anti-symmetric) superpositions correspond to even (odd) fermion parity states. (Recall that all the states $\left|\theta_{\sigma}=n\frac{\pi}{2}\right>$ are degenerate in the bulk.) Note also that these two sets of states are distinct and are related by a transformation that tranfers a single fermion between the two edges, $\Psi_s^{\dagger}\left(L\right)\Psi_s^{\phantom{\dagger}}\left(0\right) \sim e^{\pm \frac{i}{2}[\phi_{\sigma}(L) \pm \phi_{\sigma}\left(0\right)]}$. 
This is reminiscent of the four-fold ground state degeneracy present in the fully gapped time-reversal-symmetric topological phase, linked to the existence of a low energy single particle excitation at the edge of the system.
In our gapless system, however, states with a different total fermion parity also differ in their particle number, and hence are not degenerate; a system with a given particle number has only two ground states, given \emph{e.g.} by $\vert \psi_{0,+} \rangle$ and $\vert \psi_{\frac{\pi}{2},+} \rangle$.

\subsection{Symmetry fractionalization at the edges}

To better understand the nature of the topological phase, we consider the connection between time reversal and the local fermion parity at its edges. In fully gapped systems, symmetry-protected topological phases are characterized by a projective (fractionalized) representation of the symmetry operators when acting on the low-energy states of the edge~\cite{Turner2010,Turner2011,Fidkowski2011a,Chen2011,Chen2011a,Schuch2011}. In the case of a time-reversal symmetric topological superconductor, there is an anomalous relation between time reversal and fermion parity at the edge: $\mathcal{T} \mathcal{P}_{R,L} = - \mathcal{P}_{R,L} \mathcal{T}$, where $\mathcal{P}_{R,L}$ is the \emph{local} fermion parity operator acting on the right or left ends of the system, respectively~\cite{Qi2009,Chung2013}. These are local operators acting near the two edges,  defined such that in the low-energy subspace the total fermion parity operator has the form $\mathcal{P} = \mathcal{P}_L \mathcal{P}_R$. We will show that, even though our system is gapless, the gap to single fermion excitations in the bulk guarantees that a similar decomposition of fermion parity in terms of local edge operators holds; therefore, the topological phase is characterized by the same anomalous relation between time reversal and fermion parity as in the mean-field case. 

For concreteness, we demonstrate this for a system in the TS$_\mathrm{z}$ phase. 
In the low energy subspace, the total fermion parity given by Eq.~\ref{eq:FermionParity} can be written as a product of local fermion parities at the edges of the system with
\begin{equation}
\begin{split}
\mathcal{P}_{L}=i e^{-i \left(\phi_{\sigma}\left(x_1\right)-\phi_{\sigma}\left(0\right)\right)}, \\
\mathcal{P}_{R}=-i e^{-i \left(\phi_{\sigma}\left(L\right)-\phi_{\sigma}\left(x_2\right)\right)},
\end{split}
\end{equation}
where $x_{1,2}$ are arbitrary points within the bulk of the system, a few correlation lengths away from the left and right edge, respectively. We have used the fact that $\left<e^{-i \left(\phi_{\sigma}\left(x_2\right)-\phi_{\sigma}\left(x_1\right)\right)}\right>=1$, since $\phi_{\sigma}$ is pinned to a constant value in the bulk. The phase of the parity operators was chosen such that $\mathcal{P}_R^2=\mathcal{P}_L^2=1$. (Recall that at the boundary of the system $\phi_{\sigma}$ is pinned to an integer multiple of $\pi$, while in the bulk it is pinned to $\pi\left(n+\frac{1}{2}\right)$; hence $\phi_{\sigma}\left(x_1\right)-\phi_{\sigma}\left(0\right)=\phi_{\sigma}\left(L\right)-\phi_{\sigma}\left(x_2\right)=\frac{\pi}{2}\mod\pi$.)
Since the time-reversal operator $\mathcal{T}$ is anti-unitary and takes $\phi_{\sigma}$ to $-\phi_{\sigma}$, we obtain $\left\{ \mathcal{T}, \mathcal{P}_{R,L} \right\}=0$.

For a system in the trivial phase, a consistent phase choice for the parity operators that ensures $\mathcal{P}_R^2=\mathcal{P}_L^2=1$ is 
\begin{equation}
\begin{split}
\mathcal{P}_{L}=e^{-i \left(\phi_{\sigma}\left(x_1\right)-\phi_{\sigma}\left(0\right)\right)}, \\
\mathcal{P}_{R}=e^{-i \left(\phi_{\sigma}\left(L\right)-\phi_{\sigma}\left(x_2\right)\right)},
\end{split}
\end{equation}
giving the usual commutation relations $\left[\mathcal{T}, \mathcal{P}_{R,L}\right]=0$.

\section{Numerical evidence for the existence of a topological phase}

We next demonstrate the existence of the topological phase 
in a simple model, using the density
matrix renormalization group (DMRG)~\cite{White1992PRL,White1992PRB,SchollwockReview,ITensor}.

Consider spinful electrons on a 1D lattice, with spin-orbit
coupling in an alternating direction and spin-anisotropic interactions that explicitly
favor triplet pairing
\begin{equation}
\label{eq:HToyModel}
\begin{array}{ll}
H & = H_0+H_{\rm soc}+H_{\rm int} \\
H_0 & = -t\underset{j,s=\uparrow,\downarrow}{\sum}c_{j,s}^{\dagger}c^{\phantom{\dagger}}_{j+1,s}+h.c.\\
H_{\rm soc} & = iv\underset{j,s,s'=\uparrow,\downarrow}{\sum}c_{j,s}^{\dagger}\left(\frac{1+\left(-1\right)^{j}}{2}s_{s,s'}^{z}+\frac{1-\left(-1\right)^{j}}{2}s_{s,s'}^{x}\right)c^{\phantom{\dagger}}_{j+1,s'} \\ &+ h.c.\\
H_{\rm int} & = U\underset{j}{\sum}\Delta_{j}^{\dagger}\Delta^{\phantom{\dagger}}_{j},\qquad\Delta_{j}=c_{j,\uparrow}c_{j+1,\downarrow}+c_{j,\downarrow}c_{j+1,\uparrow},
\end{array}
\end{equation}
where $t$ is the hopping amplitude along the chain, $v$ is the spin-orbit
coupling strength and $U$ is the interaction strength.

In the absence of the single-particle spin-orbit coupling, i.e. for $v=0$, the $\hat{z}$
component of the spin is conserved. For any $U<0$ a spin gap opens
and we expect to find the system in the topological phase. The spin-orbit coupling term is chosen such that it breaks spin symmetry completely. We will use it to study the robustness of the topological phase without spin conservation, and its range of stability.

\begin{figure}
\subfloat[\label{fig:DMRG-ToyModel-Sz}]{\includegraphics[width=0.24\textwidth]{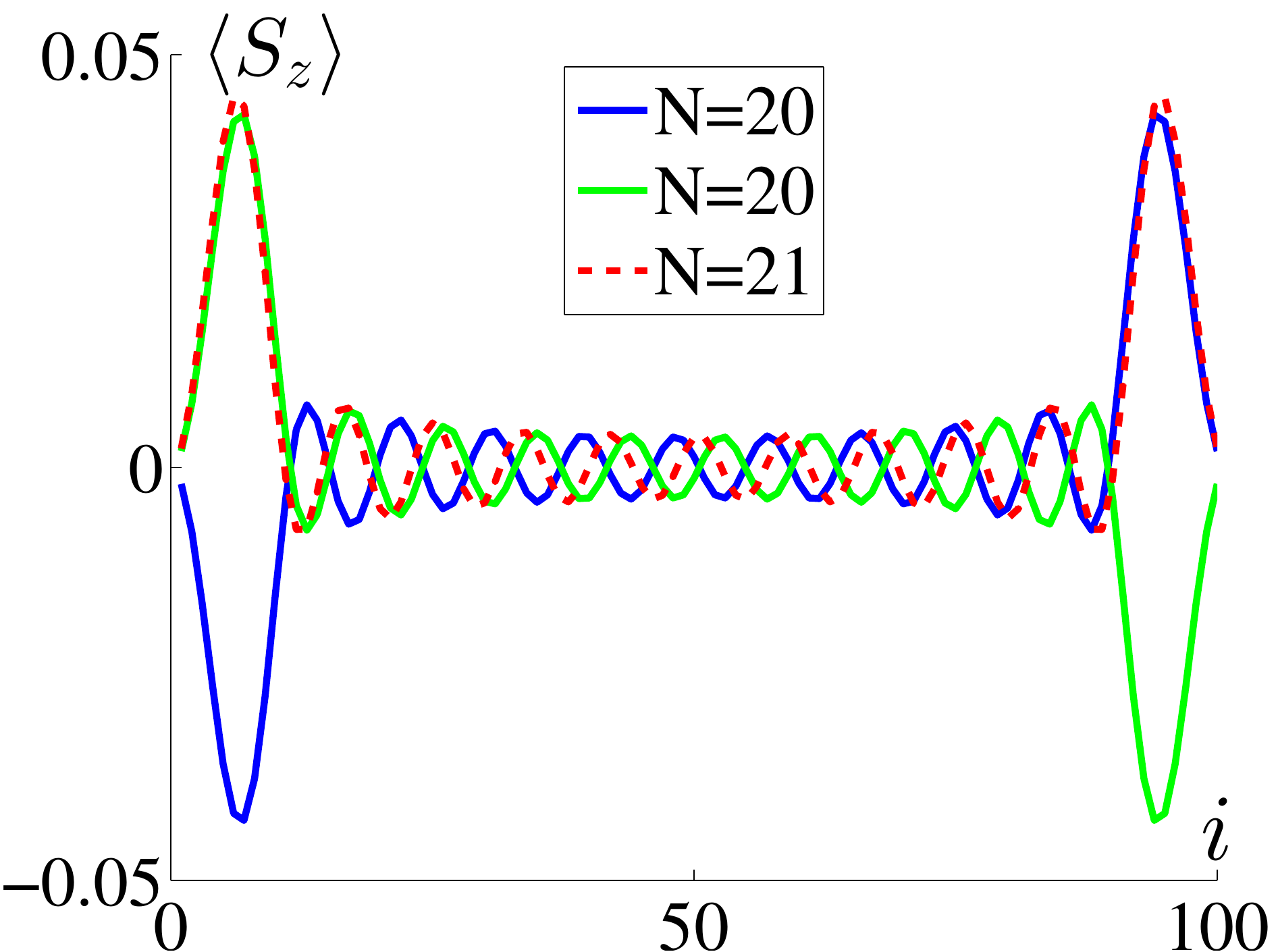}}
\subfloat[\label{fig:DMRG-ToyModel-MatrixElements}]{\includegraphics[width=0.24\textwidth]{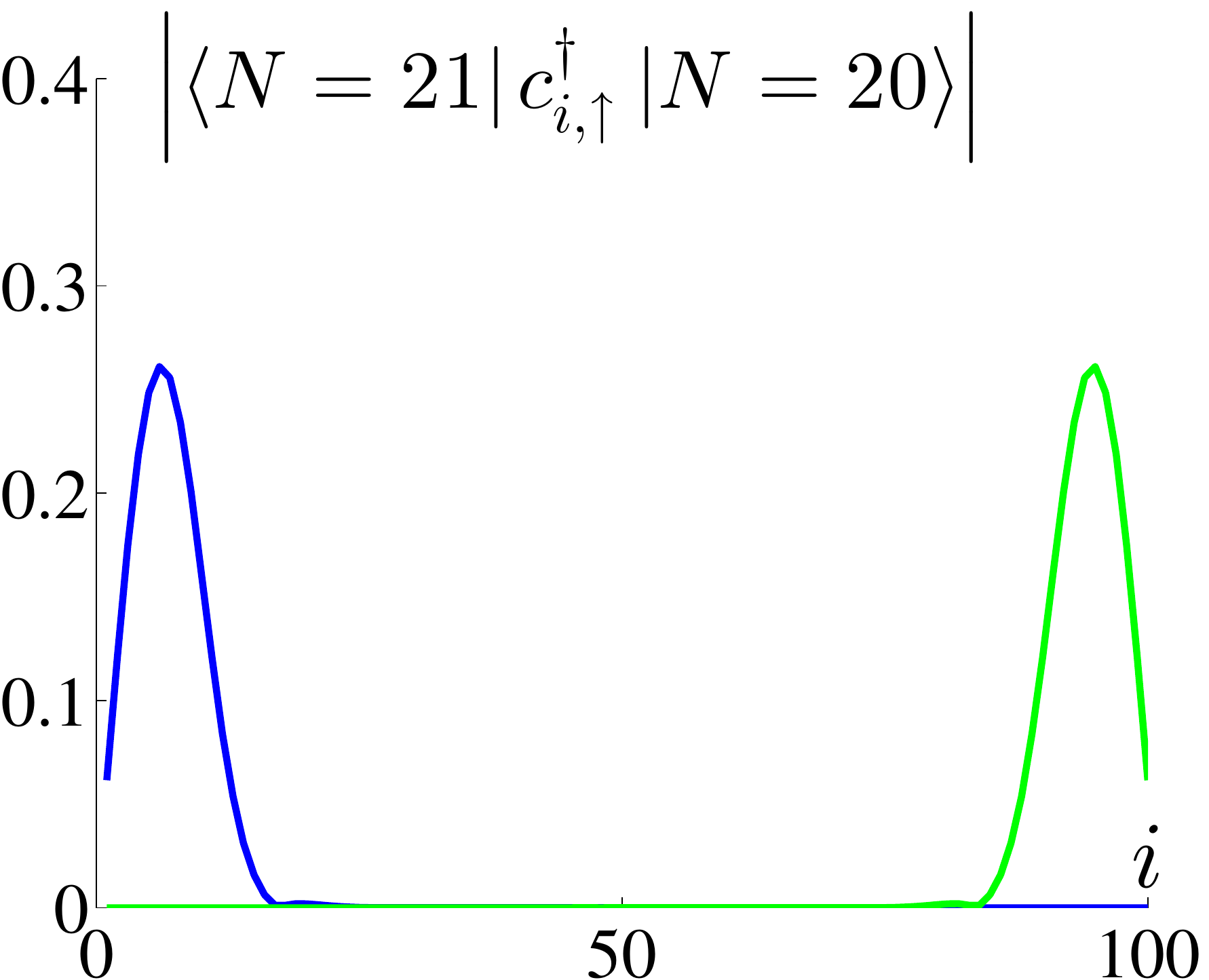}}
\caption{DMRG results for the model described by Eq. \ref{eq:HToyModel} in
absence of spin-orbit coupling, $v=0$. A system of length $L=100$
sites with model parameters $t=1,\ U=-1$ is considered. (a) Expectation
value of the $\hat{z}$ component of the spin along the chain in the
ground states of the system for even and odd number of particles.
The blue and green solid curves correspond to the two
degenerate ground states of a system with $N=20$ particles and zero
net spin. Accumulation of spin at the edge of the system is observed,
with the integrated spin in the left (right) half of the system being
$\pm\frac{1}{4}$. The red dashed curve corresponds to the ground
state of a system with $N=21$ particles and total spin $S=+\frac{1}{2}$.
Due to the non-zero spin localized at the edge, and power-law decaying
spin density wave correlations expected in a phase with $\phi_{\sigma}$ pinned to $\frac{\pi}{2}$ (see Eq. \ref{eq:OrderParamDW}), a spin density wave pattern is formed in the bulk.
(b) Matrix elements for the transition between each of the two ground
states with $N=20$ particles and the ground state with $N=21$ particles
by adding a spin up particle at a position $i$ along the lattice.
As can be seen, the matrix elements are non-zero only at either end of the system
depending on the initial state. This is in agreement with the existence of low energy single
particle states at the edges of the system.}
\end{figure}

We calculated the low energy spectrum for a system of length $L=100$ with N = 20 particles and open boundary conditions. The parameters used in this calculation are $t=1$, $U=-1$, $v=0$. The ground state is two-fold degenerate, as expected in the topological phase. We could not resolve the expected exponential
splitting between the two lowest energy states; this is presumably because the correlation length is much smaller than the system size. The two states
found by DMRG are the minimally entangled states with an integrated spin
of $\left\langle S_{z}\right\rangle =\pm\frac{1}{4}$ near the two edges.
The configuration of the $\hat{z}$ component of the spin in the two
ground states is shown in Fig.~\ref{fig:DMRG-ToyModel-Sz}. 
The topological phase is expected to have power-law decaying spin density wave correlations in the bulk (see Eq. \ref{eq:OrderParamDW} for $\phi_{\sigma}$ pinned to $\frac{\pi}{2}$). As a result, the spin polarization at the edge induces a spin density wave that decays as a power law into the system, clearly visible in the figure.

We then obtain the lowest energy state of the system with an extra spin up particle. Let us denote it by $\left|2N+1\right>$. We calculate the matrix elements between each of the states with an even number of particles, $\left|2N\right>$, to the state $\left|2N+1\right>$ by adding a spin up particle at a position $i$ along the lattice, $\left|\left\langle 2N+1\right|c_{i,\uparrow}^{\dagger}\left|2N\right\rangle \right|$ (see Fig. \ref{fig:DMRG-ToyModel-MatrixElements}). We find that the matrix elements are non-zero only at either end of the system, depending on the initial state. This is in agreement with the existence of a low energy single particle state at the edge of the system.

\begin{figure}
\includegraphics[width=0.45\textwidth]{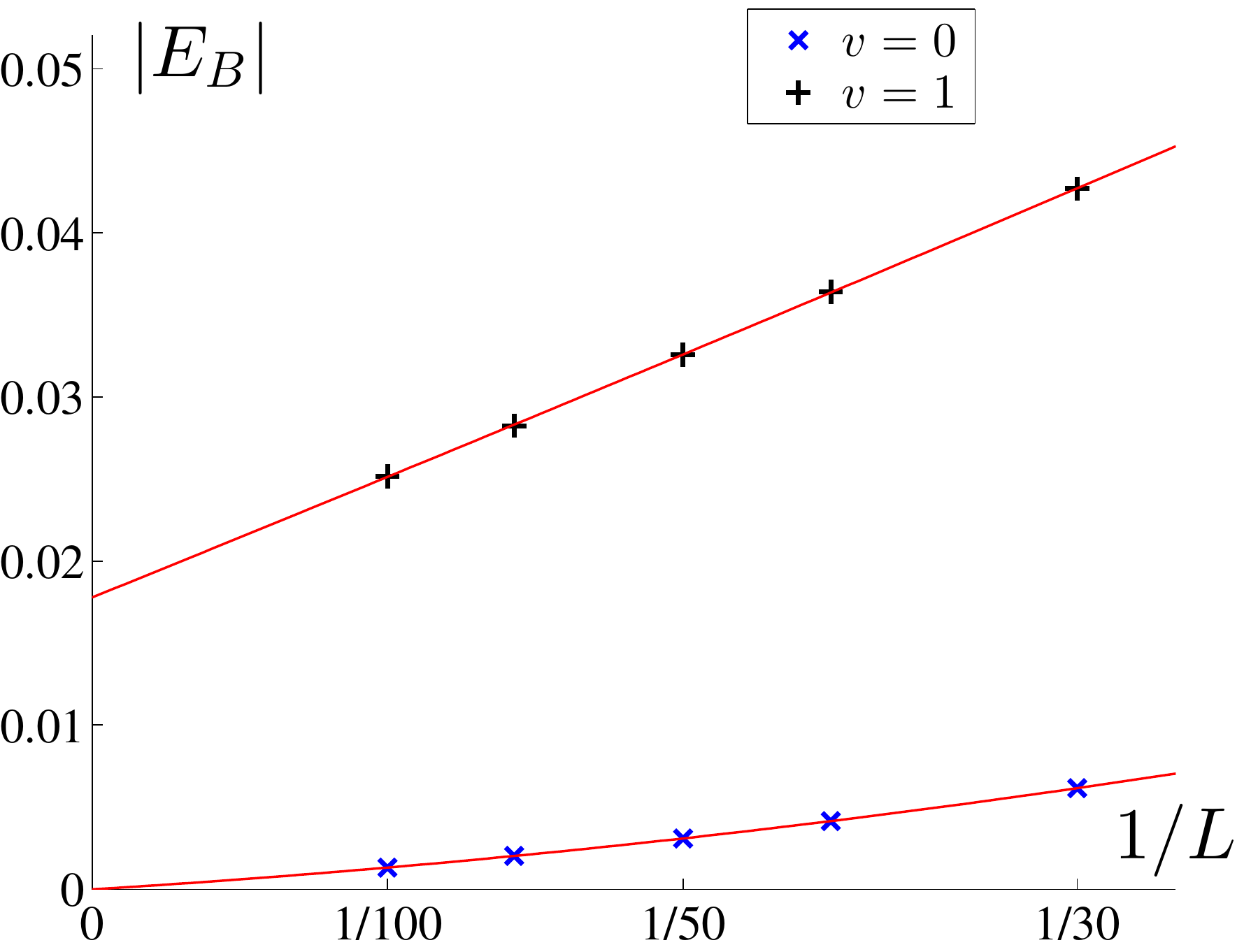}
\caption{
Pair binding energy (see Eq. \ref{eq:EBinding}) as function of system size for the model Hamiltonian given in Eq. \ref{eq:HToyModel} with parameters $t=1,\ U=-1$, spin-orbit couplings $v=0$ and $v=1$, and fixed density $n=\frac{2N}{L}=0.2$. The red solid lines are fits to parabolic curves. For $v=0$ the system is in the topological phase and the pair binding energy tends to zero as $1/L \rightarrow 0$. For $v=1$ the system is in the trivial phase and $E_B$ tends to a finite constant equal to the single particle gap in the system. }
\label{fig:DMRG-ToyModel-EB}
\end{figure}

In addition we calculated the pair binding energy (defined in Eq. \ref{eq:EBinding} above), as function of system size, keeping the density of particles fixed at $n = 0.2$ (see Fig. \ref{fig:DMRG-ToyModel-EB}). We find that it indeed tends to zero as the system becomes large, as expected in the topological phase. This once again indicates the existence of a low energy edge state, with an energy going to zero in the thermodynamic limit.

\begin{figure}
\includegraphics[width=0.45\textwidth]{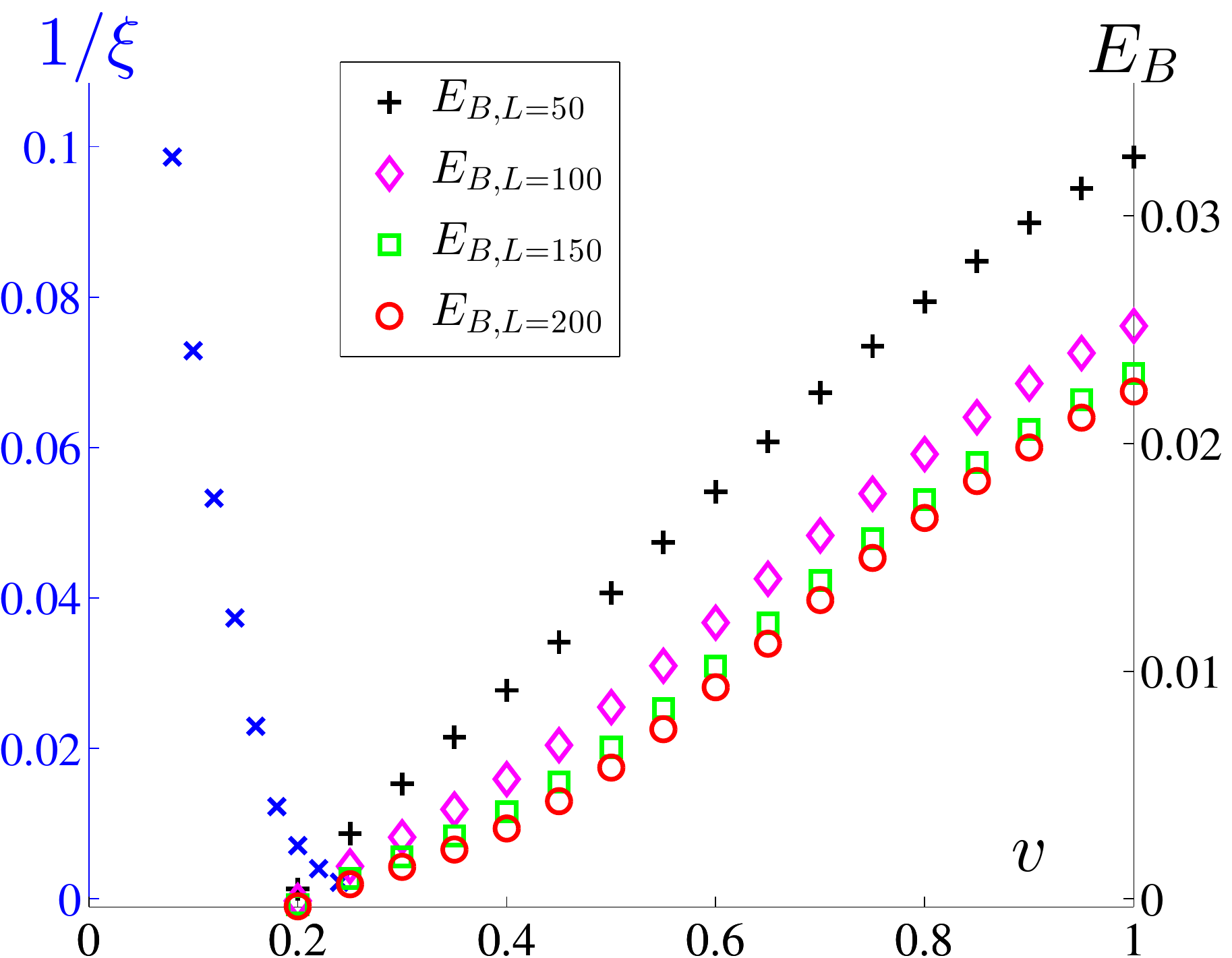}
\caption{Phase diagram obtained for the model
described by Eq. \ref{eq:HToyModel} with model parameters $t=1,\ U=-1$
and total particle density $n=\frac{N_{\uparrow}+N_{\downarrow}}{L}=0.2$.
As the spin-orbit coupling strength is increased, a phase transition
from the topological to the trivial phase is observed. For $v<v_{c}\approx0.22$
the gap (defined as the energy difference between the first excited state and the ground state) in a system with an even number of particles decreases exponentially
with the system size, as expected for the topological phase. Data
points marked by a blue cross correspond to the inverse correlation
length obtained from the finite size scaling of the gap $\Delta E = \frac{1}{L}e^{-L/\xi}$. For $v>v_{c}$
we plot the the pair binding energy, $E_B$ (see Eq. \ref{eq:EBinding}), for different system
sizes. In the limit $L\rightarrow\infty$ the pair binding energy
tends to a non-zero value, increasing with $v$, indicating an opening
of a trivial single particle gap in the system.}
\label{fig:DMRG-ToyModel-PT}
\end{figure}

For non-zero $v$, $S_{z}$ is no longer
conserved.
Moreover, conservation of $S_z$ (or any other spin component) ${\rm mod}\ 2$ is also broken, ensuring no residual symmetries are present. 
However, we find that the system remains in the topological
phase for a finite range of spin-orbit coupling strength, $v<v_{c}$.
To see this, we perform finite-size scaling of the energy gap (defined as the energy difference between the first excited state and the ground state in an even particle number sector); keeping the density constant, we choose system sizes for which
the number of particles is even. We find the gap to be exponentially
decreasing with system size, as expected in the topological phase. 
For each spin-orbit coupling strength, we extract the inverse correlation
length in the system by fitting the energy gap, $\Delta E$, vs. system size, $L$, to the form $\Delta E = \frac{1}{L}e^{-L/\xi}$. We find that the correlation length $\xi$ diverges as $v$ approaches
$v_{c}\approx 0.22$ (see Fig. \ref{fig:DMRG-ToyModel-PT}).

For $v\gtrsim v_{c}$ we calculate the pair binding energy, $E_B$, for different system sizes. As can be seen from Fig. \ref{fig:DMRG-ToyModel-PT},
the binding energy tends to a non-zero value that increases with
$v$, as the system size is increased (see also Fig. \ref{fig:DMRG-ToyModel-EB} for $E_B$ as function of system size for $v=1$). This indicates the opening of a
single particle gap and the absence of edge states in the trivial phase, as discussed in the previous section.

To emphasize the difference between the trivial and the topological phases, we calculate the ground state energy of the system, as the number of particles is varied in each phase (see Fig. \ref{fig:SpectrumVsN}). For $v=1>v_c$, the system is in the trivial phase with odd particle number states having energy larger by $\Delta$, the single particle gap, with respect to the states with an even number of particles. For $v=0<v_c$ the system is in the topological phase and there is no gap for single particle excitations; the extra particle can be accommodated at low energy at the edges.

\section{Possible realization of the topological phase - Quantum wire coupled
to a 1D superconductor}

Finally, we discuss a possible realization of the gapless topological
phase in a composite semiconducting-superconducting one-dimensional system.

\begin{figure}
\subfloat[\label{fig:CoupledWiresSetup}]
{\includegraphics[width=0.22\textwidth]{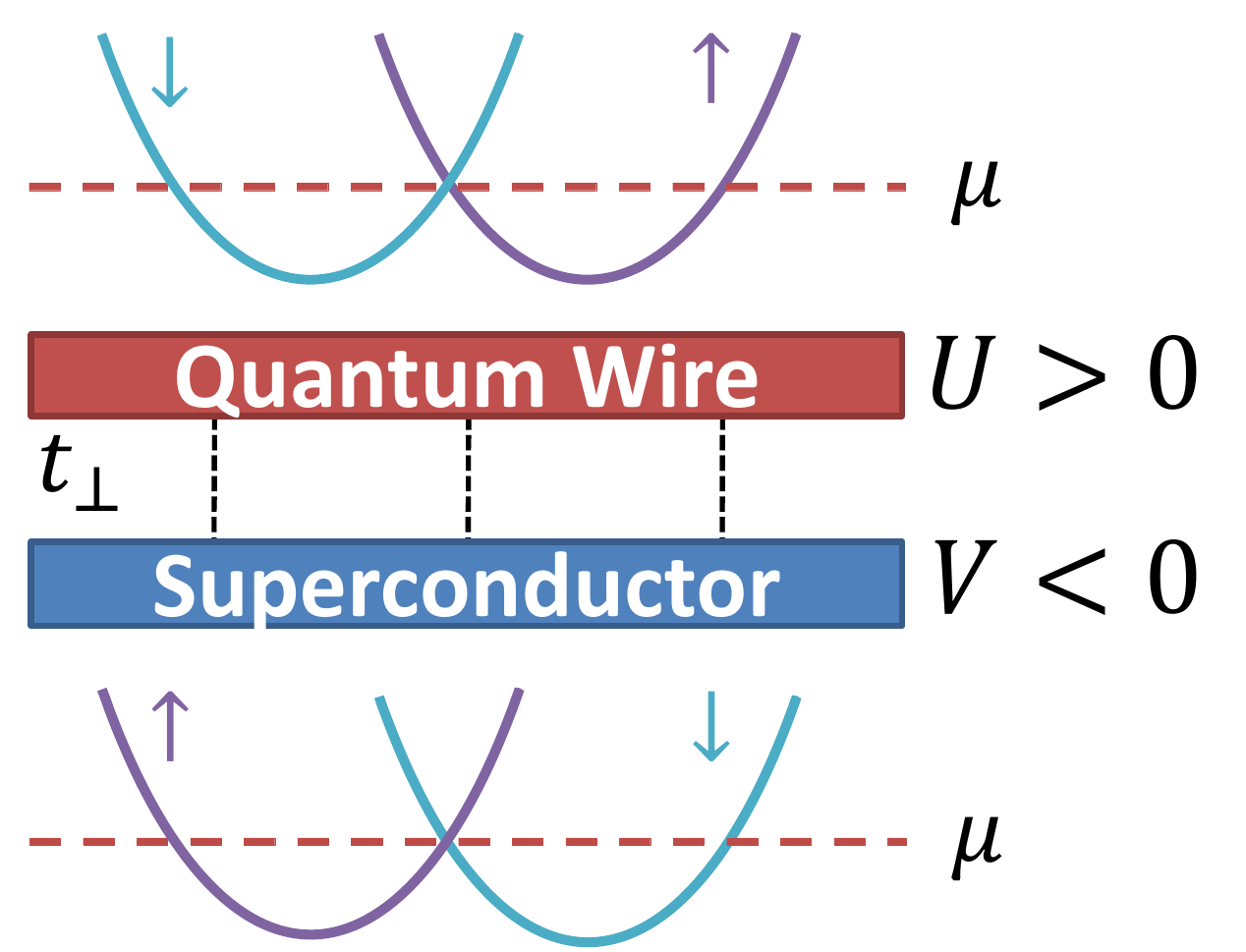}} \subfloat[\label{fig:CoupledWiresScatterings}]{\includegraphics[width=0.26\textwidth]{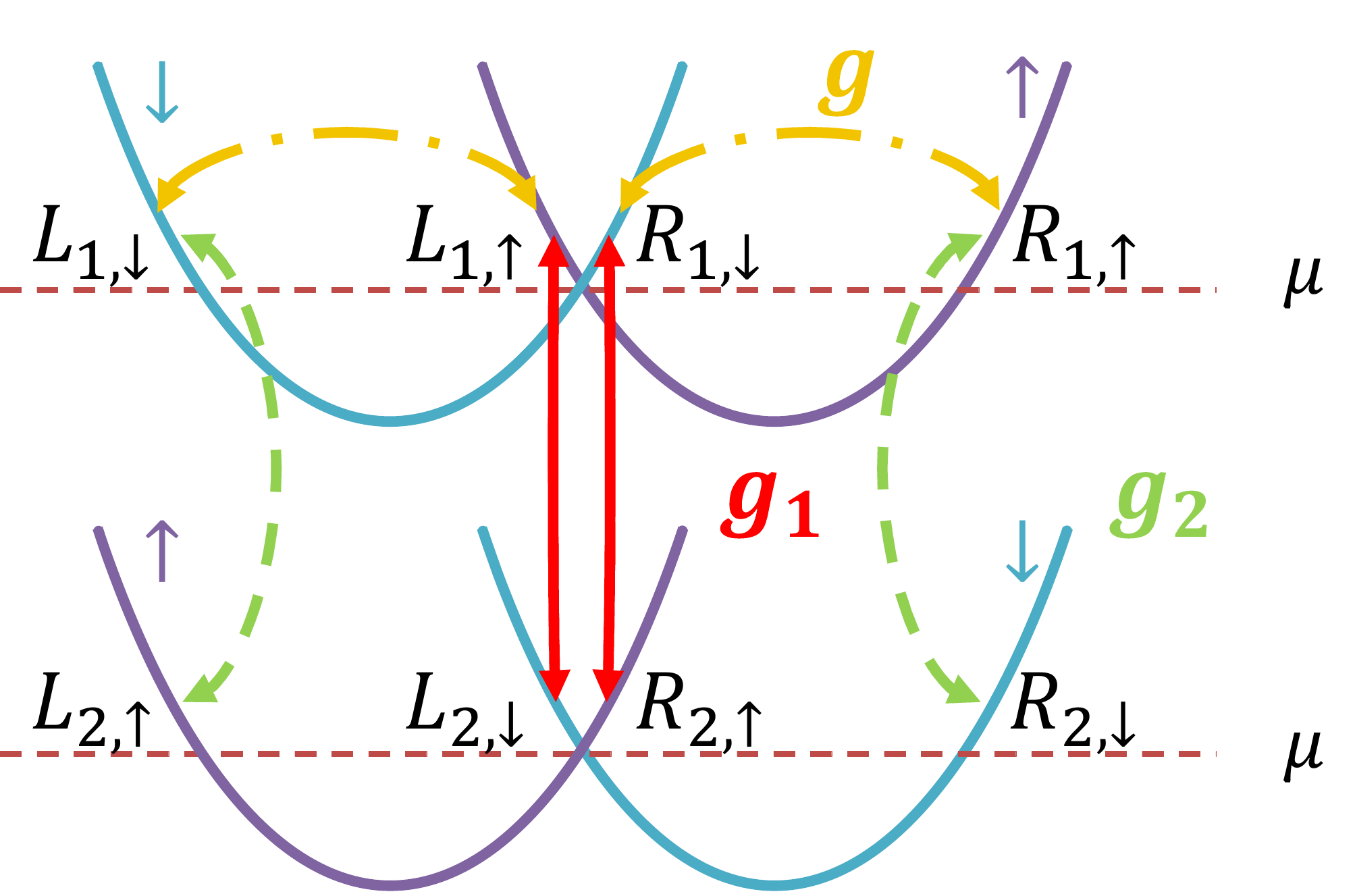}}

\caption{(a) Realization of the topological phase in a semiconducting quantum
wire with spin-orbit coupling and repulsive short-range interactions, $U>0$,
coupled to a one-dimensional superconductor with intrinsic short-range attractive
interactions $V<0$. Spin-orbit coupling is assumed to have similar magnitude and opposite sign on
the two wires, and the chemical potential is taken to be such that the
inner modes in both wires are at $k=0$ to simplify the analysis. (b) Energy dispersions in
the two wires and the relevant and marginal scattering processes.
Two independent pair tunnelings of the inner and the outer modes are
denoted by $g_{1}$ and $g_{2}$ respectively. The later is suppressed
due to the mismatch in the energy dispersions for each of the spin
flavors, as a result of the spin-orbit coupling. Back-scattering process
due to the repulsive interactions in the semiconducting wire is denoted
by $g$.}
\end{figure}

Consider a semiconducting quantum wire 
with repulsive electron-electron interactions coupled to a one-dimensional
s-wave superconductor, i.e. a wire with intrinsic attractive interactions
(see Fig.~\ref{fig:CoupledWiresSetup}). This model setup can be thought of as a one-dimensional limit of the system analyzed in~\cite{Haim2014}, where the 3D superconductor providing the pairing is replaced by a one-dimensional superconductor.

To simplify the analysis, we assume that each wire has a single transverse channel. We model the system as two one-dimensional electron gases, labelled by $i=1,2$ (the semiconducting and superconducting wires, respectively). Wire 1 has repulsive short-range density-density interactions of strength $U>0$, and wire 2 has attractive short-range interactions of strength $V<0$. Both wires have strong Rashba-type spin-orbit interactions.
As we will argue below, a particularly favourable case for realizing the topological phase is when the spin-orbit coupling terms in the two wires have a similar magnitude and an opposite sign, such that the single-particle dispersion has the structure shown in Fig.~\ref{fig:CoupledWiresSetup}, and the chemical potential is such that the inner modes in both wires are at $k=0$. We will focus on this case, and comment on the effects of deviation from it later. 
We furthermore assume that the interactions are weak, allowing for a weak-coupling renormalization group analysis. 

We linearize the modes at $k=0$ and at $k=\pm k_{F}$, denoting the
right (left) movers with spin $s$ in wire $i=1,2$ by $R\left(L\right)_{i,s}=\frac{U_{i,s}}{\sqrt{2\pi a}}e^{-i\left(\pm\frac{1}{2}\phi_{\rho,i}-\theta_{\rho,i}+s\left(\pm\frac{1}{2}\phi_{\sigma,i}-\theta_{\sigma,i}\right)\right)}$,
where $\phi_{\rho\left(\sigma\right),i}$ and $\theta_{\rho\left(\sigma\right),i}$
are the bosonic fields describing the charge (spin) modes in wire $i=1,2$. $U_{i,s}$ are Klein factors that impose the anti-commutation relation between electrons of different spins or different wires.

In absence of tunneling between the wires, attractive interactions
in the superconducting wire open a (non-topological) spin gap pinning $\phi_{\sigma,2}$
to zero. We denote the magnitude of this gap by $\Delta_{\sigma,2}$.
Repulsive interactions in the semiconducting wire give rise to a back-scattering
term $g\left(R_{1\uparrow}^{\dagger}L_{1\downarrow}^{\dagger}R^{\phantom{\dagger}}_{1\downarrow}L^{\phantom{\dagger}}_{1\uparrow}+h.c.\right)$, which in terms of the bosonic fields can be written as $\frac{g}{2\pi^2 a^2}\cos\left(2\phi_{\sigma,1}\right)$ (here $g>0$ is the backscattering coupling constant). 
As in the usual analysis of a repulsive one-dimensional electron gas, this cosine term is marginally
irrelevant and the spin sector in the semiconducting wire remains gapless.

Now consider the effect of tunneling between the wires. Due to the spin gap
in the superconducting wire, single particle tunneling between the
wires is suppressed at energies below the spin gap in the superconductor. Two independent pair tunneling (Josephson) processes are
allowed by momentum and energy conservation (see fig. \ref{fig:CoupledWiresScatterings}).
The first process describes the tunneling of a pair of electrons from the $k=0$ modes of wire 1 to the $k=0$ modes of wire 2. The corresponding term in the Hamiltonian is given by $R_{2\uparrow}^{\dagger}L_{2\downarrow}^{\dagger}R_{1\downarrow}^{\phantom{\dagger}}L_{1\uparrow}^{\phantom{\dagger}}+h.c.$. In terms of the bosonic fields it can be written as $\frac{1}{2\pi^2 a^2}\cos\left(2\theta_{\rho,-}+\phi_{\sigma,+}\right)$,
where we denote $\theta_{\rho,-}=\left(\theta_{\rho,1}-\theta_{\rho,2}\right)$
and $\phi_{\sigma,+}=\left(\phi_{\sigma,1}+\phi_{\sigma,2}\right)$.
Using second-order perturbation theory in the single-particle tunnelling amplitude $t_\perp$, we estimate the amplitude of this term as $g_{1}\sim\frac{t_{\perp}^{2}}{\Delta_{\sigma,2}}$.

Tunneling of a pair of electrons between the modes at $k=\pm k_{\rm F}$ of wire 1 to either the $k=0$ or $k=\pm k_F$ modes of wire 2 is given by $L_{2\downarrow}^{\dagger}R_{2\uparrow}^{\dagger}R_{1\uparrow}^{\phantom{\dagger}}L_{1\downarrow}^{\phantom{\dagger}}+h.c.$ or $R_{2\downarrow}^{\dagger}L_{2\uparrow}^{\dagger}R^{\phantom{\dagger}}_{1\uparrow}L^{\phantom{\dagger}}_{1\downarrow}+h.c.$; 
in terms of the bosonic fields these are given by $\frac{1}{2\pi^2 a^2}\cos\left(2\theta_{\rho,-}-\phi_{\sigma,\pm}\right)$ respectively.
These processes are suppressed with respect to the former one due to the mismatch
in the energy dispersion for each of the spin flavors between the
wires~\footnote{Here, we have assumed that the tunnelling process conserves the spin.}, originating from the different spin-orbit coupling in the two
wires. The system is therefore required to go through an intermediate excited state
at energy $\Delta_{\rm soc}\sim m\alpha^{2}$, where $\alpha$ is the
spin-orbit coupling strength. The amplitude of these scattering processes is
estimated using perturbation theory in $t_\perp$ and the interaction $V$ as $g_{2}\sim\frac{t_{\perp}^{2}V}{\Delta_{\sigma,2}\Delta_{\rm soc}}$. 

As we will see below, there is a competition between the two kinds of Josephson processes described above.
The suppression of $g_{2}$ with respect to $g_{1}$ will allow for
the formation of a topological spin gap in the semiconducting wire,
driving the system into the topological phase.

We are now ready to write the low-energy effective action for energies below the spin gap $\Delta_{\sigma,2}$ in the superconducting wire. Neglecting fluctuations of the field $\phi_{\sigma,2}$ and denoting
all the quantities in the spin sector of the semiconducting wire simply by
$\sigma$, e.g. $\phi_{\sigma,1}\rightarrow\phi_{\sigma}$, the bosonized Lagrangian density takes the form $\mathcal{L}=\mathcal{L_{\rm 0}}+\mathcal{L_{\rm int}}$, where
\begin{equation}
\begin{split}
\mathcal{L_{\rm 0}}=\underset{i=1,2}{\sum}
& \frac{1}{2\pi} K_{\rho,i} \left( \frac{1}{u_{\rho,i}} \left( \partial_{\tau} \theta_{\rho,i} \right)^2 +
u_{\rho,i} \left(\partial_x\theta_{\rho,i}\right)^2 \right) \\
& + \frac{1}{2\pi} \frac{1}{K_{\sigma}} \left( \frac{1}{u_{\sigma}} \left( \partial_{\tau} \phi_{\sigma} \right)^2 +
u_{\sigma} \left(\partial_x\phi_{\sigma}\right)^2 \right)
\end{split}
\end{equation}
is the quadratic part, with the index $i$ running over the two wires, and
\begin{equation}
\begin{split}
\mathcal{L_{\rm int}}=\frac{1}{2\pi^{2}a^{2}}&\left\{ g  \cos \left(2\phi_{\sigma}\right) \right. \\
& + \left. g_{1} \cos\left(2\theta_{\rho,-}+\phi_{\sigma}\right)+
g_{2} \cos\left(2\theta_{\rho,-}-\phi_{\sigma}\right) \right\}.
\end{split}
\end{equation}
Hereafter we denote the dimensionless couplings $\frac{g_i}{\pi a^2}$ by $y_i$.
We normalize the units of the imaginary time such that the velocity of the spin modes $u_{\sigma}$ is unity. We assume that the velocities of the charge modes on both wires are close to one and take the deviation $\delta u_{\rho,i} = u_{\rho,i} - 1$ to be a small parameter.
This allows us to rewrite $\mathcal{L_{\rm 0}}$ as
\begin{equation}
\begin{split}
\mathcal{L_{\rm 0}}=\underset{i=1,2}{\sum}
& \frac{1}{2\pi} K_{\rho,i} \left( \nabla \theta_{\rho,i} \right)^2 
+ \frac{1}{2\pi} \frac{1}{K_{\sigma}} \left( \nabla \phi_{\sigma} \right)^2 \\
& + \underset{i=1,2}{\sum} \frac{\delta u_{\rho,i}}{2\pi}  K_{\rho,i} \left[ \left( \partial_x \theta_{\rho,i} \right)^2 - \left(\partial_{\tau}\theta_{\rho,i}\right)^2 \right]
\end{split}
\end{equation}
where $\nabla$ is a 2D gradient in space-imaginary time.

We analyze the problem using weak coupling RG.
In the course of the RG flow additional couplings are generated
\begin{equation}
\mathcal{L_{\rm gen}}=\frac{1}{2\pi} \left(
K_1 \nabla \left(\theta_{\rho,1}-\theta_{\rho,2}\right) \nabla\phi_{\sigma} +
K_2 \nabla \theta_{\rho,1} \nabla \theta_{\rho,2}
\right)
\end{equation}

We perform the RG in real space using the operator product expansion (OPE) formalism~\cite{CardyBook}. At each RG step the short distance cutoff $\alpha$ is increased according to $\alpha\rightarrow\left(1+dt\right)\alpha$ while the partition function is kept fixed by renormalizing the couplings. To second order in all couplings we obtain (see Appendix~\ref{sec:Appendix-RG} for details):
\begin{align}
\begin{split}
&\frac{dK_{\sigma}^{-1}}{dt} = y^{2}+\frac{\left(y_{1}^{2}+y_{2}^{2}\right)}{4}\\
&\frac{dy}{dt} = \left(2-K_{\sigma}\right)y-\frac{y_{1}y_{2}}{2}\\
&\frac{dy_1}{dt} = \left(2-d_{y_1} + K_{1} -\frac{K_{2}}{2}\right)y_1 -\frac{yy_2}{2} \\
&\frac{dy_2}{dt} = \left(2-d_{y_2} - K_{1} -\frac{K_{2}}{2}\right)y_2 -\frac{yy_1}{2} \\
&\frac{dK_{\rho_{1,2}}}{dt} = y_{1}^{2}+y_{2}^{2}\\
&\frac{dK_{_{1}}}{dt} = y_{1}^{2}-y_{2}^{2} \\
&\frac{dK_{_{2}}}{dt} = -2\left(y_{1}^{2}+y_{2}^{2}\right),
\end{split}
\label{eq:RG}
\end{align}
where $d_{y_{1,2}}=\left(K_{\rho,1}^{-1}+K_{\rho,2}^{-1}+\frac{K_{\sigma}}{4}\right)$ is the scaling dimension of the scattering processes, and the velocities do not renormalize to this order.

Note that $y_{1,2}$ are both relevant already to first order,
as $d_{y_{1,2}}<2$ for $K_{\sigma},K_{\rho,1,2}$ close to their non-interacting value of two. Assuming initially $g_{1}\gg g_{2}$,
in agreement with the discussion above, $y_{1}$ flows to strong coupling
first. We denote by $t^{*}$ the scale at which $y_{1}$ becomes
of order unity and the perturbative analysis breaks down. The other couplings are all marginal, but begin to flow significantly as $t$ approaches $t^*$. 

Note also the competition between $y_{1,2}$ mentioned earlier, arising due to the term $-\frac{1}{2}yy_{2(1)}$ in the beta function of $y_{1(2)}$, respectively. For $y>0$ (repulsive interactions in the quantum wire), $y_1$ tends to suppress $y_2$ and change its sign (and vice versa). 
If $y_1$ and $y_2$ are initially very different in magnitude, they may end up having an opposite sign at scale $t = t^*$. Their combined contribution enhances $y$ while keeping its sign positive. This is exactly the desired situation: in order to obtain the topological phase $y$ has to be positive and relevant. Having $y_{1,2}$ with opposite signs is equivalent to inducing a superconducting gap with an opposite sign of the order parameter for the $k=0$ modes with respect to the $k=\pm k_F$ ones. For a fully gapped system, such a situation is exactly what drives the system into a (gapped) topological superconducting phase in presence of time-reversal symmetry \cite{Qi2010}.

For $t\geq t^{*},$ we assume that $\phi_{1}\equiv2\theta_{\rho,-}+\phi_{\sigma}$
becomes strongly pinned to the minimum of the $g_1$ cosine term at $\pi$. Replacing $\phi_{1}$ by its mean value, we obtain the following Lagrangian density:
\begin{align}
\begin{split}
\mathcal{L}=\frac{1}{2\pi} & \left[ \tilde{K}_{\rho,+} \left( \nabla \theta_{\rho,+} \right)^2 + \frac{1}{\tilde{K}_{\sigma}} \left( \nabla \phi_{\sigma} \right)^2 \right. \\
& \left. -\frac{1}{4}\left(K_{\rho,1}-K_{\rho,2}\right)\left(\nabla\theta_{\rho,+}\nabla\phi_{\sigma}\right) + \tilde{y} \cos\left(2\phi_{\sigma}\right) \right],
\end{split}
\label{eq:LCoupledWiresMF}
\end{align}
where we ignore the contribution of the velocities $\delta u_{\rho,i}$ as they do not change the rest of the analysis. The Luttinger parameter of the the total charge sector is given by $\tilde{K}_{\rho,+} =\frac{1}{4}\left(K_{\rho,1}+K_{\rho,2}+K_{2}\right)$. Note that this sector remains gapless as expected for a system with translational invariance. The effective parameters in the spin sector are
\begin{align}
\begin{split}
& \frac{1}{\tilde{K}_{\sigma}} = \frac{1}{K_{\sigma}}+\frac{1}{16}\left(K_{\rho,1}+K_{\rho,2}-K_{2}\right)-\frac{1}{2}K_{1} \\
& \tilde{y} = y-y_{2}.
\end{split}
\end{align}
The cross term $\nabla\theta_{\rho,+}\nabla\phi_{\sigma}$ in the Lagrangian (\ref{eq:LCoupledWiresMF}) changes the scaling dimensions of the cosine term. However, treating it perturbatively we find that it does not change the RG equations to
second order. The RG flow in the spin sector for $t>t^{*}$ then takes the standard Kosterlitz-Thouless
form
\begin{align}
\begin{split}
& \frac{d}{dt}\left(\tilde{K}_{\sigma}^{-1}\right)=\tilde{y}^{2} \\
& \frac{d\tilde{y}}{dt}=\left(2-\tilde{K}_{\sigma}\right)\tilde{y}
\end{split}
\label{eq:MFParams}
\end{align}
The topological phase is then obtained if $\tilde{y}\left(t^{*}\right)$
is positive and the cosine term is relevant, i.e. if
$\frac{1}{2}\tilde{K}_{\sigma}\left(t^{*}\right)<\left(\frac{1+\tilde{y}\left(t^{*}\right)}{1-\tilde{y}\left(t^{*}\right)}\right)^{1/2}$. If the cosine term is irrelevant, the spin sector remains gapless.

\begin{figure}
\includegraphics[width=0.45\textwidth]{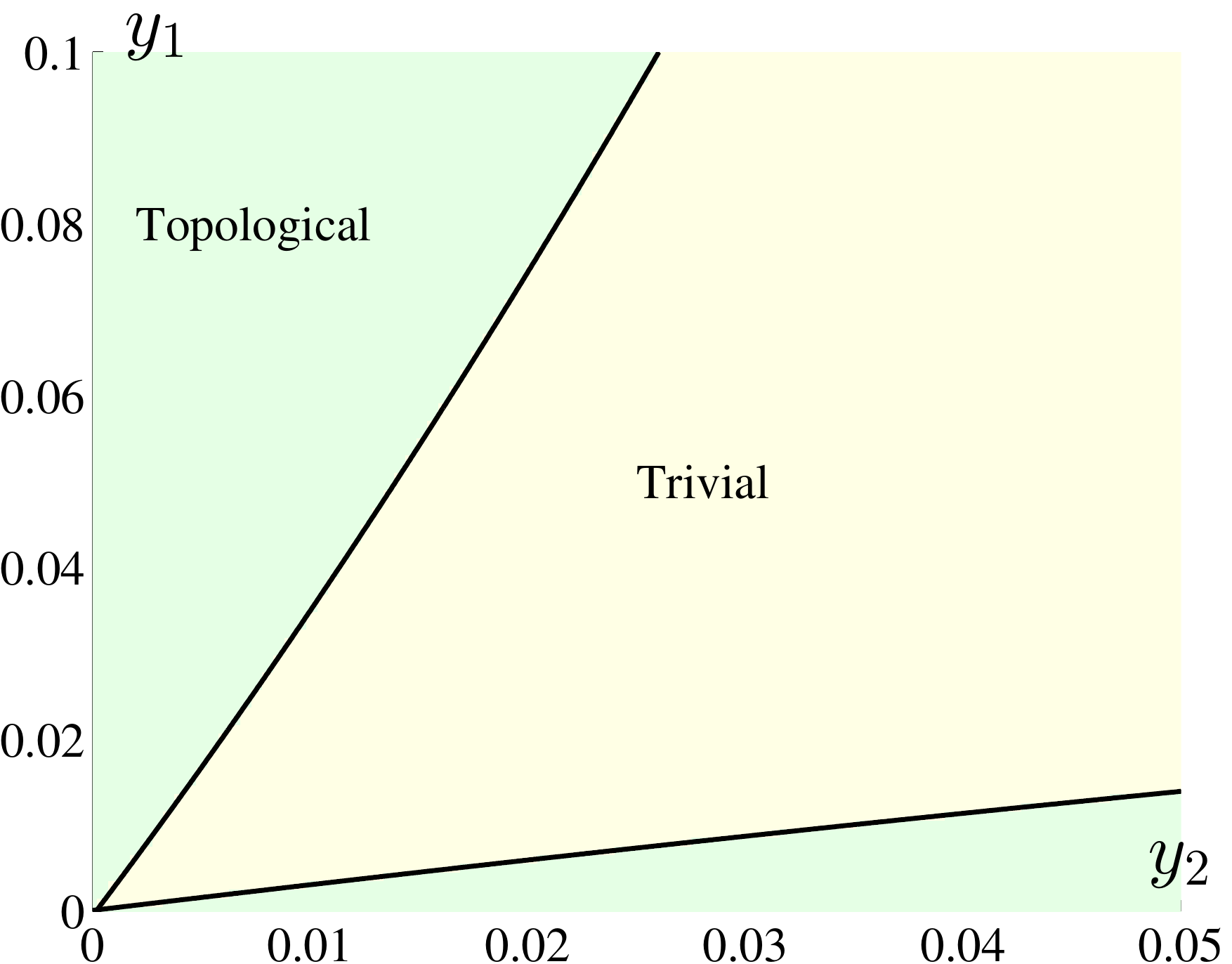}
\caption{Phase diagram for the composite semiconducting-superconducting one-dimensional system,
as function of the bare dimensionless pair tunneling couplings $y_{1,2}\left(t=0\right)$.
Initial conditions corresponding to spin-isotropic interactions in the quantum wire are used, $\frac{1}{2}K_{\sigma}\left(t=0\right)=1+y\left(t=0\right)$, where the strength of the repulsive interactions in the wire is taken to be $y\left(t=0\right)=0.1$. We assume the non-interacting values for the charge sectors in each wire $K_{\rho_{1,2}}\left(t=0\right)=2$. The scale $t^*$ at which the integration of the RG equations (\ref{eq:RG}) is stopped and the field $\phi_1$ is replaced by its mean value 
was defined such that $y_1(t^*) = 0.5$.
As argued in the text, increasing the spin-orbit coupling strength in the wires suppresses
$y_{2}$ with respect to $y_{1}$. As can be seen from the phase diagram this drives the system into the topological phase. The topological phase is obtained also for $y_{2}\gg y_{1}$ as all the analysis is
symmetric in $y_{1,2}$.}
\label{fig:CoupledWiresPD}
\end{figure}

We can now obtain the phase diagram of the system as a function of the bare values of the pair tunnelings $y_{1,2}\left(t=0\right)$ as follows.
The RG equations (\ref{eq:RG}) are integrated up to the scale $t^*$ at which $y_{1}$ reaches a value of order unity. At this scale, we replace $\phi_1$ by its mean value and obtain the effective action (\ref{eq:LCoupledWiresMF}). The flow continues according to Eq.~\ref{eq:MFParams}. If the cosine term is relevant,
the resulting phase is inferred from the sign of $\tilde{y}\left(t^{*}\right)$: $\tilde{y}(t^*)>0$ ($\tilde{y}(t^*)<0$) corresponds to the topological (trivial) phase, respectively.

The resulting phase diagram is shown in Fig.~\ref{fig:CoupledWiresPD}. 
We used initial conditions that correspond to spin-isotropic interactions in the quantum wire, $\frac{1}{2}K_{\sigma}\left(t=0\right)=1+y\left(t=0\right)$. We set $y\left(t=0\right)$ to $0.1$.
For simplicity we assume the non-interacting values for the charge sectors in each wire $K_{\rho_{1,2}}\left(t=0\right)=2$. The scale $t^*$ was defined such that $y_1(t^*) = 0.5$. Neither of these choices changes the phase diagram qualitatively.

We find that the cosine term always flows to strong coupling opening
a spin gap in the semiconducting wire. The topological phase is obtained
for $y_{1}$ sufficiently larger than $y_{2}$, corresponding to pair tunneling of the outer
modes $g_{2}$ being suppressed with respect to the pair tunneling
of the inner modes $g_{1}$.
In our microscopic system this situation can be achieved
due to the form of the spin-orbit coupling in the wires. Setting the chemical potential to the point indicated in Fig.~\ref{fig:CoupledWiresScatterings} is optimal for this purpose;  
large enough spin-orbit coupling will eventually drive
the system into the topological phase, since it suppresses $g_2$ but not $g_1$. Note that the topological phase
can also be obtained for $y_{2}\gg y_{1}$, as all the analysis is
symmetric in $y_{1,2}$.

\section{Discussion}

We have shown that a gapless topological phase, with exponentially localized edge states, can exist in a \emph{strictly} one-dimensional time-reversal symmetric (class DIII) system with a conserved number of particles. The localization of the edge states arises due to a gap to single particle excitations in the bulk of the system. The edge states are characterized by an anomalous relation between the fermion parity and time reversal symmetry: time reversal flips the local fermion parity at the two edges, just as in the case of a DIII mean-field topological superconductor~\cite{Qi2009}.

It is interesting to note that gapless topologically protected phases are possible in other symmetry classes in one dimension, as well (see, e.g., Refs.~\cite{Kestner2011,Iemini2015}). In particular, class BDI (with particle-hole symmetry and time reversal that squares to $+1$) is likely to support similar phases, although there are less distinct phases than in the non-interacting case~\cite{Fidkowski2010}. 
As we have shown here, as long as there is a gap for single fermion excitations in the bulk \emph{even in the absence of long-range order}, the corresponding topological phase may be robust in a gapless one-dimensional system. 

We discussed a possible realization of the phase by proximity coupling a semiconducting wire with spin-orbit coupling to a superconducting wire. Another realization of the phase may be possible in quasi-one-dimensional organic conductors where spin anisotropic interactions are believed to be present. As was analyzed in Ref.~\cite{GiamarchiSchulz1986}, these interactions could also drive the system into a phase with dominant triplet superconducting correlations, that we identified as a topological phase. In such a system, inter-chain hopping eventually drives the system into a three-dimensional long-range ordered superconducting phase; however, if the system is very anisotropic, the properties of the topological phase (e.g., the presence of zero-energy surface states) may already be apparent at intermediate temperatures above the critical temperature, where the system is described as nearly-decoupled fluctuating one-dimensional superconductors. 

It is important to note that, unlike the fully gapped topological superconducting phase which is immune to non-magnetic disorder (as long as its strength does not exceed the superconducting gap), the gapless phase can be sensitive even to weak disorder. 
In the topological gapless phase $2k_F$ density fluctuations are gapped. Therefore the dominant effect of an impurity potential is due to the $4k_F$ density fluctuations term.
The effect of this term depends strongly on the nature of interactions in the system, parametrized by the Luttinger parameter of the charge sector, $K_{\rho}$. For $K_{\rho}$ sufficiently large (in the conventions used here, $K_\rho>1$, where $K_\rho = 2$ corresponds to a non-interacting system), backscattering due to a weak point-like impurity potential flows to zero under RG, leaving the topological phase intact. If the interactions are too strongly repulsive, corresponding to $K_{\rho}<1$, any weak impurity potential becomes relevant. In the presence of many such impurities, the system becomes localized; this destroys the topological nature of the phase, and gaps out the edge states. 

We conclude that the topological phase is robust in the presence of weak disorder, as long as the repulsive interactions in the system are not too strong. A metallic gate placed near the system may be used to screen the long-range part of the Coulomb interactions, thus making the topological phase more stable to disorder. 



\section{Acknowledgments}

We would like to thank Liang Fu, Arbel Haim, Yuval Oreg, Yoni Schattner,  Ady Stern, Yochai Werman, and Konrad W\"{o}lms for fruitful discussions. We are particularly indebted to Miles Stoudenmire for his help with setting up the DMRG calculations using the ITensor package. This research was supported by the Minerva foundation, a Minerva ARCHES prize, a Marie Curie CIG grant, a GIF-Young Researcher grant, and by the Israel Science Foundation. 

\appendix

\section{Tunneling density of states at an edge of a gapless topological system}
\label{sec:Appendix-TDOS}
In this section we describe the calculation of the matrix element given
in Eq. \ref{eq:MatrixElement}, which contributes to the $\omega\rightarrow0$ tunneling
density of states (TDOS) for a system in the gapless topological phase.

Consider a topological region that extends from $x=0$ to $x=L$, with
a trivial region on its either side. The bosonized Hamiltonian describing
the system is given by
\begin{eqnarray}
H&=&\int dx\underset{\alpha=\rho,\sigma}{\sum}\frac{u_{\alpha}}{2\pi}\left[K_{\alpha}\left(\partial_{x}\theta_{\alpha}\right)^{2}+\frac{1}{K_{\alpha}}\left(\partial_{x}\phi_{\alpha}\right)^{2}\right] \nonumber \\
&+&\int dx\frac{g\left(x\right)}{2\pi^{2}a^{2}}\cos\left(2\phi_{\sigma}\right),
\end{eqnarray}
where $g\left(x\right)=g>0$ in the topological region $0<x<L$, and
$g\left(x\right)=g_{0}<0$ in the trivial regions, $x\leq0$ and $x\geq L$.
We assume the couplings $g,g_{0}$ are large enough such that $\phi_{\sigma}$
is pinned close to the minimum of the potential in each region. This
allows us to expand the cosine around $\pi n_{1,2}$ in the trivial
regions and around $\pi\left(m+\frac{1}{2}\right)$ in the topological
region, where $n_{1,2},m$ are integers. 
We consider a variational Hamiltonian where the cosine term is replaced with a quadratic potential, consistent with such an expansion. There are two distinct configurations of the potential, corresponding to different states:
\begin{equation*}
V_{0}\left(x\right)=\frac{1}{2\pi^{2}a^{2}}\begin{cases}
-2g_{0}\phi_{\sigma}^{2} & x<0,\ x>L\\
2g\left(\phi_{\sigma}-\frac{\pi}{2}\right)^{2} & 0<x<L
\end{cases}
\end{equation*}
and
\begin{equation*}
V_{1}\left(x\right)=\frac{1}{2\pi^{2}a^{2}}\begin{cases}
-2g_{0}\left(\phi_{\sigma}-\pi\right)^{2} & x<0\\
2g\left(\phi_{\sigma}-\frac{\pi}{2}\right)^{2} & 0<x<L\\
-2g_{0}\phi_{\sigma}^{2} & x>L
\end{cases}.
\end{equation*}
The respective variational Hamiltonians are denoted by $H_{0,1}$ and their ground
states by $\left|0\right\rangle $ and $\left|1\right\rangle $. Saddle
point configurations of $\phi_{\sigma}$ in these states, $\phi_{\sigma,0,1}\left(x\right)$,
are depicted in Fig. \ref{fig:4foldDegen}. We denote the difference between them by
$\Delta\phi_{\sigma}=\phi_{\sigma,1}-\phi_{\sigma,0}$. Note that
the energy of the spin sector for these two configurations is the
same, and they are therefore nearly degenerate. The difference in
energy arrises due to the charging energy (which scales as $\frac{1}{L}$,
tending to zero for large enough system), as in the state $\left|1\right\rangle $
the total spin in the system is $\frac{1}{2}$, corresponding to an
odd number of particles, while in the state $\left|0\right\rangle $
the total number of particles is even. Since the density of the particles
in the system is given by $-\frac{1}{\pi}\partial_x\phi_{\rho}$, the
respective saddle point configurations of $\phi_{\rho}$ in these
two states differ by a constant gradient term $\Delta\phi_{\rho}=\pi\left(1-\frac{x}{L}\right)$
in the region $0<x<L$.
The two ground states are therefore related by a unitary transformation
$\left|1\right\rangle =e^{-i\hat{\delta}}\left|0\right\rangle $, where
$\hat{\delta}=\underset{\alpha=\rho,\sigma}{\sum}\hat{\delta}_{\alpha}$ and
\begin{equation*}
\hat{\delta}_{\alpha}=\frac{1}{\pi}\int_{-\infty}^{\infty}dx'\Delta\phi_{\alpha}\left(x'\right)\partial_{x'}\theta_{\alpha}\left(x'\right).
\end{equation*}
To see this note that under this transformation the field $\phi_{\alpha}\rightarrow\phi_{\alpha}+\Delta\phi_{\alpha}$.

The matrix element that is expected to give the largest contribution
to the TDOS at $\omega\rightarrow0$ can now be written as
\begin{equation*}
\left\langle 1\right|\Psi_{\uparrow}^{\dagger}\left(x\right)\left|0\right\rangle =\left\langle e^{i\hat{\delta}}\Psi_{\uparrow}^{\dagger}\left(x\right)\right\rangle,
\end{equation*}
where an expectation value with respect to the ground state of $H_{0}$
is assumed in the final expression.
Writing the single particle creation
operator as
\begin{equation*}
\Psi_{\uparrow}^{\dagger}\left(x\right)\sim\underset{r}{\sum}e^{-irk_{F}x}e^{i\phi_{r,\uparrow}},
\end{equation*}
where $\phi_{r,\uparrow}=\frac{1}{2}r\phi_{\rho}\left(x\right)-\theta_{\rho}\left(x\right)+\frac{1}{2}r\phi_{\sigma}\left(x\right)-\theta_{\sigma}\left(x\right)$
and the sum over $r=\pm1$ stands for right and left movers respectively,
the matrix element becomes
\begin{equation*}
\left\langle e^{i\hat{\delta}}\Psi_{\uparrow}^{\dagger}\left(x\right)\right\rangle \sim\underset{r}{\sum}e^{-irk_{F}x}\left\langle e^{i\hat{\delta}}e^{i\phi_{r,\uparrow}}\right\rangle.
\end{equation*}
To proceed we therefore need to diagonalize $H_{0}$ and find the
expansion of the bosonic fields $\phi_{\rho,\sigma}$ and $\theta_{\rho,\sigma}$
in terms of its eigenmodes. Note that since the spin and charge sectors
are decoupled, the expectation value above can be written as a product
\begin{equation*}
\left\langle e^{i\hat{\delta}}e^{i\phi_{r,\downarrow}}\right\rangle =\underset{\alpha=\rho,\sigma}{\prod}\left\langle e^{i\hat{\delta}_{\alpha}}e^{i\left(\frac{1}{2}r\phi_{\alpha}\left(x\right)-\theta_{\alpha}\left(x\right)\right)}\right\rangle _{\alpha},
\end{equation*}
where $\left\langle ... \right\rangle _{\alpha=\rho,\sigma}$ denotes
the expectation value in the ground state of the charge and spin sectors
respectively.

We begin by diagonalizing the spin sector, $H_{0,\sigma}$. For simplicity
we take the spin gap in the trivial regions to be infinite, $\left|g_{0}\right|\rightarrow\infty$
(equivalently we can assume that the topological region is surrounded by vacuum).
This pins the field $\phi_{\sigma}$ and the current $\partial_{x}\theta_{\sigma}$
at the boundary to zero, i.e. $\left.\phi_{\sigma}\right|_{x=0,L}=0$
and $\left.\partial_{x}\theta_{\sigma}\right|_{x=0,L}=0$. A mode
expansion for the fields $\phi_{\sigma},\theta_{\sigma}$ then takes
the following form:
\begin{equation*}
\begin{split}
\theta_{\sigma}\left(x\right)&=i\overset{\infty}{\underset{k=1}{\sum}}\sqrt{\frac{1}{K_{\sigma}k}}\cos\left(\frac{\pi kx}{L}\right)\left(a_{k}^{\phantom{\dagger}}-a_{k}^{\dagger}\right) \\
\phi_{\sigma}\left(x\right)&=\phi_{\sigma,0}\left(x\right)+\overset{\infty}{\underset{k=1}{\sum}}\sqrt{\frac{K_{\sigma}}{k}}\sin\left(\frac{\pi kx}{L}\right)\left(a_{k}^{\phantom{\dagger}}+a_{k}^{\dagger}\right)
\end{split}
\end{equation*}
where the expansion for $\phi_{\sigma}$ is around the constant saddle
point solution $\phi_{\sigma,0}\left(x\right)$, and $a_{k}^{\dagger},a_{k}^{\phantom{\dagger}}$
are bosonic creation and annihilation operators satisfying the commutation
relations $\left[a_{k}^{\phantom{\dagger}},a_{k'}^{\dagger}\right]=\delta_{k,k'}$.

The Hamiltonian translates into
\begin{equation*}
H_{0,\sigma}=\frac{1}{2}\overset{\infty}{\underset{k=1}{\sum}}\left[A_{k}\left(a_{k}^{\phantom{\dagger}}a_{k}^{\dagger}+a_{k}^{\dagger}a_{k}^{\phantom{\dagger}}\right)+B_{k}\left(a_{k}^{2\phantom{\dagger}}+a_{k}^{\dagger2}\right)\right]
\end{equation*}
where $A_{k}=u_{\sigma}\frac{\pi k}{L}+\frac{\left|g\right|K_{\sigma}}{\pi a^{2}}\frac{L}{\pi k}$,
$B_{k}=\frac{\left|g\right|K_{\sigma}}{\pi a^{2}}\frac{L}{\pi k}$,
and a Bogoliubov transformation $a_{k}^{\phantom{\dagger}}=\alpha_{k}b_{k}^{\phantom{\dagger}}+\beta_{k}b_{k}^{\dagger}$
can be used to bring it into a diagonal form
\begin{equation*}
H_{0,\sigma}=\sum E^{\vphantom{\dagger}}_{k}b_{k}^{\dagger}b_{k}^{\phantom{\dagger}}+const.
\end{equation*}
The coefficients in the transformation are $\alpha_{k}=\sqrt{\frac{1}{2}\left(\frac{A_{k}}{E_{k}}+1\right)}$,
$\beta_{k}=-\sqrt{\frac{1}{2}\left(\frac{A_{k}}{E_{k}}-1\right)}$.
The eigen-energies are given by $E_{k}=\sqrt{A_{k}^{2}-B_{k}^{2}}$ and
$b_{k}^{\dagger},b_{k}^{\phantom{\dagger}}$ are the bosonic creation and annihilation
operators of the corresponding eigen-modes.

The fields $\phi_{\sigma},\theta_{\sigma}$ are given by linear combinations
of these creation and annihilation operators. Hence, the expectation
value can be calculated using the identity
\begin{equation*}
\left\langle e^{i\hat{\delta}_{\sigma}}e^{i\left(\frac{1}{2}r\phi_{\sigma}\left(x\right)-\theta_{\sigma}\left(x\right)\right)}\right\rangle _{\sigma}=e^{-\frac{1}{2}\left\langle \left(\hat{\delta}_{\sigma}+\frac{1}{2}r\phi_{\sigma}\left(x\right)-\theta_{\sigma}\left(x\right)\right)^{2}\right\rangle_{\sigma} }.
\end{equation*}
Denote the spin gap in the system by $\Delta_{\sigma}=\sqrt{\frac{2\left|g\right|u_{\sigma}K_{\sigma}}{\pi a^{2}}}$
and the inverse correlation length by $\kappa=\xi^{-1}=\Delta_{\sigma}/u_{\sigma}$.
In the limit $\left|g_{0}\right|\rightarrow\infty$
the difference between the saddle point configurations is
\begin{equation*}
\Delta\phi_{\sigma}=\begin{cases}
\pi & x<0\\
\pi \frac{\sinh[\kappa(L-x)]}{\sinh(\kappa L)} & 0<x<L\\
0 & x>L
\end{cases}.
\end{equation*}
We can then write $\hat{\delta}_{\sigma}=\theta_{\sigma}\left(0\right)+\Delta\theta_{\sigma}$,
where $\Delta\theta_{\sigma}=\int_{0}^{L}dxe^{-\kappa x}\partial_{x}\theta_{\sigma}\left(x\right)$ for $\kappa L\gg1$. Note also that we are interested only in the absolute value of the
matrix element. The expectation values $\left\langle \phi_{\sigma}\left(x\right)\theta_{\sigma}\left(x\right)\right\rangle $
and $\left\langle \phi_{\sigma}\left(x\right)\hat{\delta}_{\sigma}\right\rangle $
are imaginary and only contribute a constant
phase shift to the oscillations in $2k_{F}$ of the matrix element squared. The asymptotic behavior
is dictated by the function
\begin{equation*}
F_{\sigma}\equiv\left\langle \left(\theta_{\sigma}\left(x\right)-\hat{\delta}_{\sigma}\right)^{2}\right\rangle +\frac{1}{4}\left\langle \phi_{\sigma}^{2}\left(x\right)\right\rangle.
\end{equation*}
Denoting $\gamma_{k}=\left(1+\kappa^{2}/\left(\frac{\pi k}{L}\right)^{2}\right)^{1/2}$
we obtain
\begin{multline*}
F_{\sigma}=\frac{1}{K_{\sigma}}\overset{\infty}{\underset{k=1}{\sum}}\frac{\gamma_{k}}{k}\left[\cos\left(\frac{\pi kx}{L}\right)-\frac{\kappa^{2}}{\kappa^{2}+\left(\frac{\pi k}{L}\right)^{2}}\right]^{2}\\
+ \frac{1}{4}K_{\sigma}\overset{\infty}{\underset{k=1}{\sum}}\frac{1}{k\gamma_{k}}\sin^{2}\left(\frac{\pi kx}{L}\right).
\end{multline*}
Taking the continuum limit $\frac{1}{L}\rightarrow0$, the sum over
$k$ can be written as an integral over $q=\frac{\pi k}{L}$. Performing the integral and considering the asymptotics for $\xi\ll x\ll L$ we find that
\begin{equation*}
F_{\sigma}\sim\frac{\pi}{2}\frac{1}{K_{\sigma}}\kappa x.
\end{equation*}

A similar calculation for the charge sector gives the following asymptotic behavior for $a\ll x\ll L$
\begin{multline*}
F_{\rho}\equiv\left\langle \left(\theta_{\rho}\left(x\right)-\hat{\delta}_{\rho}\right)^{2}\right\rangle +\frac{1}{4}\left\langle \phi_{\rho}^{2}\left(x\right)\right\rangle \\
\sim \frac{1}{2}\left(\frac{1}{K_{\rho}}+\frac{1}{4}K_{\rho}\right)\ln\left(\frac{x}{a}\right)+\frac{1}{K_{\rho}}\ln\left(\frac{L}{a}\right).
\end{multline*}

The absolute value of the matrix element squared is therefore
\begin{multline*}
\left|\left\langle e^{i\hat{\delta}}\Psi_{\uparrow}^{\dagger}\left(x\right)\right\rangle \right|^{2}\sim e^{-\left(F_{\sigma}+F_{\rho}\right)} \\
\sim\left(\frac{a}{L}\right)^{\frac{1}{K_{\rho}}}\left(\frac{a}{x}\right)^{\frac{\alpha}{2}}e^{-\frac{\pi}{2K_{\sigma}}\frac{x}{\xi}},
\end{multline*}
where we denote $\alpha=\frac{1}{K_{\rho}}+\frac{1}{4}K_{\rho}$ and
$\xi=\frac{u_{\sigma}}{\Delta_{\sigma}}$ is the correlation length.

\section{RG flow equations obtained using the operator product expansion}
\label{sec:Appendix-RG}
In this appendix we give further details on the derivation of the
RG flow equations~(\ref{eq:RG}). As mentioned in the main text, we perform the
RG in real space. At each step we rescale the short distance cutoff $\alpha$,
defined as the minimal distance between two operators in the theory,
according to $\alpha\rightarrow\left(1+dt\right)\alpha$. We use the operator product
expansion (OPE) to replace pairs of operators which are within the
new short distance cutoff, allowing us to recover the original action
with renormalized couplings.

The general form for an OPE of two operators $O_{i}$, $O_{j}$ is
\begin{equation*}
:O_{i}\left(r_{1}\right)::O_{j}\left(r_{2}\right):=\underset{k}{\sum}C_{ijk}\left(r_{1}-r_{2}\right):O_{k}\left(\frac{r_{1}+r_{2}}{2}\right):,
\end{equation*}
where $:O_{i}:$ denotes normal ordering of the operator. The equality
is valid only when both sides of the equation are considered
inside a correlation function with another operator (or set of operators)
at a distance much greater than $\left|r_{1}-r_{2}\right|$ from the
operators $O_{i,j}$. The functions $C_{ijk}\left(r_{1}-r_{2}\right)$
have the form $C_{ijk}=\frac{c_{ijk}}{\left|r_{1}-r_{2}\right|^{d_{i}+d_{j}-d_{k}}}$, where $d_{i,j,k}$ are the scaling dimensions of the corresponding
operators and $c_{ijk}$ are the OPE coefficients which are pure numbers.
The beta functions to second order of the respective couplings $g_{i,j,k}$
are then given by~\cite{CardyBook}
\begin{equation*}
\frac{dg_{k}}{dt}=\left(d-d_{k}\right)g_{k}-\underset{i,j}{\sum}c_{ijk}g_{i}g_{j},
\end{equation*}
where $d$ is the dimension of the problem.

We demonstrate the OPE explicitly for the pair of operators $O_{1}=\nabla\theta_{\rho,-}\nabla\phi_{\sigma}$ and $O_{2}=\cos\left(2\theta_{\rho,-}+\phi_{\sigma}\right)$, both of which we treat perturbatively.
We start by writing the cosine as a sum of exponents, and expending the exponent as a series
\begin{multline*}
:e^{i\left(2\theta_{\rho,-}+\phi_{\sigma}\right)}:=
\underset{n}{\sum}\frac{i^{n}}{n!}:\left(2\theta_{\rho,-}+\phi_{\sigma}\right)^{n}:= \\
\underset{n}{\sum}\frac{i^{n}}{n!}\underset{m}{\sum}{n \choose m}2^m:\theta^m_{\rho,-}\phi_{\sigma}^{n-m}:
\end{multline*}
The normal ordered product of the operators $:O_1\left(r_1\right)::O_2\left(r_2\right):$ will then contain a summation over terms of the form
\begin{equation*}
:\partial_{i}\theta_{\rho,-}\left(r_{1}\right)\partial_{i}\phi_{\sigma}\left(r_{1}\right)::\theta_{\rho,-}^{m}\left(r_{2}\right)\phi_{\sigma}^{n-m}\left(r_{2}\right):.
\end{equation*}
In each such term, we can contract $\partial_{i}\theta_{\rho,-}\left(r_{1}\right)$ with one of the $\theta_{\rho,-}\left(r_{2}\right)$, giving $m$ possible configurations. For each of them, there are $n-m$ ways to contract $\partial_{i}\phi_{\sigma}\left(r_{1}\right)$ with one of the $\phi_{\sigma}\left(r_{2}\right)$. These contractions are given by
\begin{equation*}
\left\langle \partial_{i}\phi_{\sigma}\left(r_{1}\right)\phi_{\sigma}\left(r_{2}\right)\right\rangle =-\frac{1}{2}K_{\sigma}\frac{\left(r_{1}-r_{2}\right)_{i}}{\left|r_{1}-r_{2}\right|^{2}}
\end{equation*}
and
\begin{multline*}
\left\langle \partial_{i}\theta_{\rho,-}\left(r_{1}\right)\theta_{\rho,-}\left(r_{2}\right)\right\rangle = \\
\left\langle \partial_{i}\theta_{\rho,1}\left(r_{1}\right)\theta_{\rho,1}\left(r_{2}\right)\right\rangle +\left\langle \partial_{i}\theta_{\rho,2}\left(r_{1}\right)\theta_{\rho,2}\left(r_{2}\right)\right\rangle = \\
-\frac{1}{2}\left(K_{\rho1}^{-1}+K_{\rho2}^{-1}\right)\frac{\left(r_{1}-r_{2}\right)_{i}}{\left|r_{1}-r_{2}\right|^{2}}
\end{multline*}
respectively.
The product of the operators now becomes
\begin{widetext}
\begin{equation*}
:\nabla\theta_{\rho,-}\nabla\phi_{\sigma}::e^{i\left(2\theta_{\rho,-}+\phi_{\sigma}\right)}:=\frac{1}{4}K_{\sigma}\left(K_{\rho,1}^{-1}+K_{\rho,2}^{-1}\right)\frac{1}{\left|r_{1}-r_{2}\right|^{2}}\underset{n}{\sum}\frac{i^{n}}{n!}\underset{m}{\sum}{n \choose m} 2^m m\left(n-m\right):\theta_{\rho,-}^{m-1}\phi_{\sigma}^{n-m-1}:
\end{equation*}
\end{widetext}
Rewriting the sum over $m$ as
\begin{multline*}
2n\left(n-1\right)\underset{m}{\sum} {n-2 \choose m-1}
:\left(2\theta_{\rho,-}\right)^{m-1}\phi_{\sigma}^{n-2-\left(m-1\right)}:= \\
2n\left(n-1\right):\left(2\theta_{\rho,-}+\phi_{\sigma}\right)^{n-2}:
\end{multline*}
and the sum over $n$ as
\begin{equation*}
2i^{2}\underset{n}{\sum}\frac{i^{n-2}}{\left(n-2\right)!}:\left(2\theta_{\rho,-}+\phi_{\sigma}\right)^{n-2}:=-2:e^{i\left(2\theta_{\rho,-}+\phi_{\sigma}\right)}:
\end{equation*}
we obtain
\begin{multline*}
:\nabla\theta_{\rho,-}\nabla\phi_{\sigma}::\cos
\left(2\theta_{\rho,-}+\phi_{\sigma}\right):= \\ -\frac{1}{2}K_{\sigma}\left(K_{\rho,1}^{-1}+K_{\rho,2}^{-1}\right)\frac{1}{\left|r_{1}-r_{2}\right|^{2}}:\cos\left(2\theta_{\rho,-}+\phi_{\sigma}\right):,
\end{multline*}
where we identify $-\frac{1}{2}K_{\sigma}\left(K_{\rho,1}^{-1}+K_{\rho,2}^{-1}\right)$
as the OPE coefficient. This results in the following contribution to the beta function of
$y_1$
\begin{equation*}
\frac{dy_1}{dt}=\frac{1}{2}K_{\sigma}\left(K_{\rho,1}^{-1}+K_{\rho,2}^{-1}\right)K_{1}y_{1}.
\end{equation*}
To second order in the coupling constants, we can use the non-interacting values for the Luttinger parameters $K_{\sigma}=K_{\rho,1,2}=2$
and write
\begin{equation*}
\frac{dy_1}{dt}=K_1 y_1.
\end{equation*}

\bibliography{ref}

\begin{thebibliography}{51}%
\makeatletter
\providecommand \@ifxundefined [1]{%
 \@ifx{#1\undefined}
}%
\providecommand \@ifnum [1]{%
 \ifnum #1\expandafter \@firstoftwo
 \else \expandafter \@secondoftwo
 \fi
}%
\providecommand \@ifx [1]{%
 \ifx #1\expandafter \@firstoftwo
 \else \expandafter \@secondoftwo
 \fi
}%
\providecommand \natexlab [1]{#1}%
\providecommand \enquote  [1]{``#1''}%
\providecommand \bibnamefont  [1]{#1}%
\providecommand \bibfnamefont [1]{#1}%
\providecommand \citenamefont [1]{#1}%
\providecommand \href@noop [0]{\@secondoftwo}%
\providecommand \href [0]{\begingroup \@sanitize@url \@href}%
\providecommand \@href[1]{\@@startlink{#1}\@@href}%
\providecommand \@@href[1]{\endgroup#1\@@endlink}%
\providecommand \@sanitize@url [0]{\catcode `\\12\catcode `\$12\catcode
  `\&12\catcode `\#12\catcode `\^12\catcode `\_12\catcode `\%12\relax}%
\providecommand \@@startlink[1]{}%
\providecommand \@@endlink[0]{}%
\providecommand \url  [0]{\begingroup\@sanitize@url \@url }%
\providecommand \@url [1]{\endgroup\@href {#1}{\urlprefix }}%
\providecommand \urlprefix  [0]{URL }%
\providecommand \Eprint [0]{\href }%
\providecommand \doibase [0]{http://dx.doi.org/}%
\providecommand \selectlanguage [0]{\@gobble}%
\providecommand \bibinfo  [0]{\@secondoftwo}%
\providecommand \bibfield  [0]{\@secondoftwo}%
\providecommand \translation [1]{[#1]}%
\providecommand \BibitemOpen [0]{}%
\providecommand \bibitemStop [0]{}%
\providecommand \bibitemNoStop [0]{.\EOS\space}%
\providecommand \EOS [0]{\spacefactor3000\relax}%
\providecommand \BibitemShut  [1]{\csname bibitem#1\endcsname}%
\let\auto@bib@innerbib\@empty
\bibitem [{Bon()}]{Bonderson2013}%
  \BibitemOpen
  \href@noop {} {}\bibinfo {note} {For a general discussion of this question,
  see: P. Bonderson and C. Nayak, Physical Review B \textbf{87}, 195451
  (2013).}\BibitemShut {Stop}%
\bibitem [{\citenamefont {Read}\ and\ \citenamefont
  {Green}(2000)}]{ReadGreen2000}%
  \BibitemOpen
  \bibfield  {author} {\bibinfo {author} {\bibfnamefont {N.}~\bibnamefont
  {Read}}\ and\ \bibinfo {author} {\bibfnamefont {D.}~\bibnamefont {Green}},\
  }\href {\doibase 10.1103/PhysRevB.61.10267} {\bibfield  {journal} {\bibinfo
  {journal} {Phys. Rev. B}\ }\textbf {\bibinfo {volume} {61}},\ \bibinfo
  {pages} {10267} (\bibinfo {year} {2000})}\BibitemShut {NoStop}%
\bibitem [{\citenamefont {Kitaev}(2009)}]{kitaev2009periodic}%
  \BibitemOpen
  \bibfield  {author} {\bibinfo {author} {\bibfnamefont {A.}~\bibnamefont
  {Kitaev}},\ }in\ \href {http://dx.doi.org/10.1063/1.3149495} {\emph {\bibinfo
  {booktitle} {AIP Conf. Proc.}}},\ Vol.\ \bibinfo {volume} {1134}\ (\bibinfo
  {year} {2009})\ p.~\bibinfo {pages} {22}\BibitemShut {NoStop}%
\bibitem [{\citenamefont {Ryu}\ \emph {et~al.}(2010)\citenamefont {Ryu},
  \citenamefont {Schnyder}, \citenamefont {Furusaki},\ and\ \citenamefont
  {Ludwig}}]{Ryu2010}%
  \BibitemOpen
  \bibfield  {author} {\bibinfo {author} {\bibfnamefont {S.}~\bibnamefont
  {Ryu}}, \bibinfo {author} {\bibfnamefont {A.~P.}\ \bibnamefont {Schnyder}},
  \bibinfo {author} {\bibfnamefont {A.}~\bibnamefont {Furusaki}}, \ and\
  \bibinfo {author} {\bibfnamefont {A.~W.~W.}\ \bibnamefont {Ludwig}},\ }\href
  {\doibase 10.1088/1367-2630/12/6/065010} {\bibfield  {journal} {\bibinfo
  {journal} {New Journal of Physics}\ }\textbf {\bibinfo {volume} {12}},\
  \bibinfo {pages} {065010} (\bibinfo {year} {2010})}\BibitemShut {NoStop}%
\bibitem [{\citenamefont {Volovik}(2009)}]{volovik2009universe}%
  \BibitemOpen
  \bibfield  {author} {\bibinfo {author} {\bibfnamefont {G.~E.}\ \bibnamefont
  {Volovik}},\ }\href@noop {} {\emph {\bibinfo {title} {The universe in a
  helium droplet}}}\ (\bibinfo  {publisher} {Oxford University Press},\
  \bibinfo {year} {2009})\BibitemShut {NoStop}%
\bibitem [{\citenamefont {Hasan}\ and\ \citenamefont
  {Kane}(2010)}]{HasanKane2010}%
  \BibitemOpen
  \bibfield  {author} {\bibinfo {author} {\bibfnamefont {M.}~\bibnamefont
  {Hasan}}\ and\ \bibinfo {author} {\bibfnamefont {C.}~\bibnamefont {Kane}},\
  }\href {\doibase 10.1103/RevModPhys.82.3045} {\bibfield  {journal} {\bibinfo
  {journal} {Reviews of Modern Physics}\ }\textbf {\bibinfo {volume} {82}},\
  \bibinfo {pages} {3045} (\bibinfo {year} {2010})}\BibitemShut {NoStop}%
\bibitem [{\citenamefont {Qi}\ and\ \citenamefont {Zhang}(2011)}]{QiZhang2011}%
  \BibitemOpen
  \bibfield  {author} {\bibinfo {author} {\bibfnamefont {X.-L.}\ \bibnamefont
  {Qi}}\ and\ \bibinfo {author} {\bibfnamefont {S.-C.}\ \bibnamefont {Zhang}},\
  }\href {\doibase 10.1103/RevModPhys.83.1057} {\bibfield  {journal} {\bibinfo
  {journal} {Reviews of Modern Physics}\ }\textbf {\bibinfo {volume} {83}},\
  \bibinfo {pages} {1057} (\bibinfo {year} {2011})}\BibitemShut {NoStop}%
\bibitem [{\citenamefont {Kitaev}(2001)}]{Kitaev2001}%
  \BibitemOpen
  \bibfield  {author} {\bibinfo {author} {\bibfnamefont {A.~Y.}\ \bibnamefont
  {Kitaev}},\ }\href {\doibase 10.1070/1063-7869/44/10S/S29} {\bibfield
  {journal} {\bibinfo  {journal} {Physics-Uspekhi}\ }\textbf {\bibinfo {volume}
  {44}},\ \bibinfo {pages} {131} (\bibinfo {year} {2001})}\BibitemShut
  {NoStop}%
\bibitem [{\citenamefont {Oreg}\ \emph {et~al.}(2010)\citenamefont {Oreg},
  \citenamefont {Refael},\ and\ \citenamefont {von Oppen}}]{Oreg2010}%
  \BibitemOpen
  \bibfield  {author} {\bibinfo {author} {\bibfnamefont {Y.}~\bibnamefont
  {Oreg}}, \bibinfo {author} {\bibfnamefont {G.}~\bibnamefont {Refael}}, \ and\
  \bibinfo {author} {\bibfnamefont {F.}~\bibnamefont {von Oppen}},\ }\href
  {\doibase 10.1103/PhysRevLett.105.177002} {\bibfield  {journal} {\bibinfo
  {journal} {Phys. Rev. Lett.}\ }\textbf {\bibinfo {volume} {105}},\ \bibinfo
  {pages} {177002} (\bibinfo {year} {2010})}\BibitemShut {NoStop}%
\bibitem [{\citenamefont {Lutchyn}\ \emph {et~al.}(2010)\citenamefont
  {Lutchyn}, \citenamefont {Sau},\ and\ \citenamefont
  {Das~Sarma}}]{Lutchyn2010}%
  \BibitemOpen
  \bibfield  {author} {\bibinfo {author} {\bibfnamefont {R.~M.}\ \bibnamefont
  {Lutchyn}}, \bibinfo {author} {\bibfnamefont {J.~D.}\ \bibnamefont {Sau}}, \
  and\ \bibinfo {author} {\bibfnamefont {S.}~\bibnamefont {Das~Sarma}},\ }\href
  {\doibase 10.1103/PhysRevLett.105.077001} {\bibfield  {journal} {\bibinfo
  {journal} {Phys. Rev. Lett.}\ }\textbf {\bibinfo {volume} {105}},\ \bibinfo
  {pages} {077001} (\bibinfo {year} {2010})}\BibitemShut {NoStop}%
\bibitem [{\citenamefont {Alicea}(2012)}]{Alicea2012}%
  \BibitemOpen
  \bibfield  {author} {\bibinfo {author} {\bibfnamefont {J.}~\bibnamefont
  {Alicea}},\ }\href {\doibase 10.1088/0034-4885/75/7/076501} {\bibfield
  {journal} {\bibinfo  {journal} {Rep. Prog. Phys.}\ }\textbf {\bibinfo
  {volume} {75}},\ \bibinfo {pages} {076501} (\bibinfo {year}
  {2012})}\BibitemShut {NoStop}%
\bibitem [{\citenamefont {Beenakker}(2013)}]{Beenakker2013}%
  \BibitemOpen
  \bibfield  {author} {\bibinfo {author} {\bibfnamefont {C.}~\bibnamefont
  {Beenakker}},\ }\href {\doibase 10.1146/annurev-conmatphys-030212-184337}
  {\bibfield  {journal} {\bibinfo  {journal} {Annual Review of Condensed Matter
  Physics}\ }\textbf {\bibinfo {volume} {4}},\ \bibinfo {pages} {113} (\bibinfo
  {year} {2013})}\BibitemShut {NoStop}%
\bibitem [{\citenamefont {Fidkowski}\ \emph {et~al.}(2011)\citenamefont
  {Fidkowski}, \citenamefont {Lutchyn}, \citenamefont {Nayak},\ and\
  \citenamefont {Fisher}}]{Fidkowski2011}%
  \BibitemOpen
  \bibfield  {author} {\bibinfo {author} {\bibfnamefont {L.}~\bibnamefont
  {Fidkowski}}, \bibinfo {author} {\bibfnamefont {R.}~\bibnamefont {Lutchyn}},
  \bibinfo {author} {\bibfnamefont {C.}~\bibnamefont {Nayak}}, \ and\ \bibinfo
  {author} {\bibfnamefont {M.}~\bibnamefont {Fisher}},\ }\href {\doibase
  10.1103/PhysRevB.84.195436} {\bibfield  {journal} {\bibinfo  {journal} {Phys.
  Rev. B}\ }\textbf {\bibinfo {volume} {84}},\ \bibinfo {pages} {195436}
  (\bibinfo {year} {2011})}\BibitemShut {NoStop}%
\bibitem [{\citenamefont {Sau}\ \emph {et~al.}(2011)\citenamefont {Sau},
  \citenamefont {Halperin}, \citenamefont {Flensberg},\ and\ \citenamefont
  {Das~Sarma}}]{Sau2011}%
  \BibitemOpen
  \bibfield  {author} {\bibinfo {author} {\bibfnamefont {J.}~\bibnamefont
  {Sau}}, \bibinfo {author} {\bibfnamefont {B.}~\bibnamefont {Halperin}},
  \bibinfo {author} {\bibfnamefont {K.}~\bibnamefont {Flensberg}}, \ and\
  \bibinfo {author} {\bibfnamefont {S.}~\bibnamefont {Das~Sarma}},\ }\href
  {\doibase 10.1103/PhysRevB.84.144509} {\bibfield  {journal} {\bibinfo
  {journal} {Phys. Rev. B}\ }\textbf {\bibinfo {volume} {84}},\ \bibinfo
  {pages} {144509} (\bibinfo {year} {2011})}\BibitemShut {NoStop}%
\bibitem [{\citenamefont {Ruhman}\ \emph {et~al.}(2014)\citenamefont {Ruhman},
  \citenamefont {Berg},\ and\ \citenamefont {Altman}}]{ruhman2014topological}%
  \BibitemOpen
  \bibfield  {author} {\bibinfo {author} {\bibfnamefont {J.}~\bibnamefont
  {Ruhman}}, \bibinfo {author} {\bibfnamefont {E.}~\bibnamefont {Berg}}, \ and\
  \bibinfo {author} {\bibfnamefont {E.}~\bibnamefont {Altman}},\ }\href@noop {}
  {\bibfield  {journal} {\bibinfo  {journal} {arXiv preprint arXiv:1412.3444}\
  } (\bibinfo {year} {2014})}\BibitemShut {NoStop}%
\bibitem [{\citenamefont {Qi}\ \emph {et~al.}(2009)\citenamefont {Qi},
  \citenamefont {Hughes}, \citenamefont {Raghu},\ and\ \citenamefont
  {Zhang}}]{Qi2009}%
  \BibitemOpen
  \bibfield  {author} {\bibinfo {author} {\bibfnamefont {X.-L.}\ \bibnamefont
  {Qi}}, \bibinfo {author} {\bibfnamefont {T.~L.}\ \bibnamefont {Hughes}},
  \bibinfo {author} {\bibfnamefont {S.}~\bibnamefont {Raghu}}, \ and\ \bibinfo
  {author} {\bibfnamefont {S.-C.}\ \bibnamefont {Zhang}},\ }\href {\doibase
  10.1103/PhysRevLett.102.187001} {\bibfield  {journal} {\bibinfo  {journal}
  {Phys. Rev. Lett.}\ }\textbf {\bibinfo {volume} {102}},\ \bibinfo {pages}
  {187001} (\bibinfo {year} {2009})}\BibitemShut {NoStop}%
\bibitem [{\citenamefont {Fu}\ and\ \citenamefont {Berg}(2010)}]{Fu2010}%
  \BibitemOpen
  \bibfield  {author} {\bibinfo {author} {\bibfnamefont {L.}~\bibnamefont
  {Fu}}\ and\ \bibinfo {author} {\bibfnamefont {E.}~\bibnamefont {Berg}},\
  }\href {\doibase 10.1103/PhysRevLett.105.097001} {\bibfield  {journal}
  {\bibinfo  {journal} {Phys. Rev. Lett.}\ }\textbf {\bibinfo {volume} {105}},\
  \bibinfo {pages} {097001} (\bibinfo {year} {2010})}\BibitemShut {NoStop}%
\bibitem [{\citenamefont {Deng}\ \emph {et~al.}(2012)\citenamefont {Deng},
  \citenamefont {Viola},\ and\ \citenamefont {Ortiz}}]{Deng2012}%
  \BibitemOpen
  \bibfield  {author} {\bibinfo {author} {\bibfnamefont {S.}~\bibnamefont
  {Deng}}, \bibinfo {author} {\bibfnamefont {L.}~\bibnamefont {Viola}}, \ and\
  \bibinfo {author} {\bibfnamefont {G.}~\bibnamefont {Ortiz}},\ }\href
  {\doibase 10.1103/PhysRevLett.108.036803} {\bibfield  {journal} {\bibinfo
  {journal} {Phys. Rev. Lett.}\ }\textbf {\bibinfo {volume} {108}},\ \bibinfo
  {pages} {036803} (\bibinfo {year} {2012})}\BibitemShut {NoStop}%
\bibitem [{\citenamefont {Nakosai}\ \emph {et~al.}(2012)\citenamefont
  {Nakosai}, \citenamefont {Tanaka},\ and\ \citenamefont
  {Nagaosa}}]{Nakosai2012}%
  \BibitemOpen
  \bibfield  {author} {\bibinfo {author} {\bibfnamefont {S.}~\bibnamefont
  {Nakosai}}, \bibinfo {author} {\bibfnamefont {Y.}~\bibnamefont {Tanaka}}, \
  and\ \bibinfo {author} {\bibfnamefont {N.}~\bibnamefont {Nagaosa}},\ }\href
  {\doibase 10.1103/PhysRevLett.108.147003} {\bibfield  {journal} {\bibinfo
  {journal} {Phys. Rev. Lett.}\ }\textbf {\bibinfo {volume} {108}},\ \bibinfo
  {pages} {147003} (\bibinfo {year} {2012})}\BibitemShut {NoStop}%
\bibitem [{\citenamefont {Seradjeh}(2012)}]{Seradjeh2012}%
  \BibitemOpen
  \bibfield  {author} {\bibinfo {author} {\bibfnamefont {B.}~\bibnamefont
  {Seradjeh}},\ }\href {\doibase 10.1103/PhysRevB.86.121101} {\bibfield
  {journal} {\bibinfo  {journal} {Phys. Rev. B}\ }\textbf {\bibinfo {volume}
  {86}},\ \bibinfo {pages} {121101(R)} (\bibinfo {year} {2012})}\BibitemShut
  {NoStop}%
\bibitem [{\citenamefont {Wong}\ and\ \citenamefont {Law}(2012)}]{Wong2012}%
  \BibitemOpen
  \bibfield  {author} {\bibinfo {author} {\bibfnamefont {C.~L.~M.}\
  \bibnamefont {Wong}}\ and\ \bibinfo {author} {\bibfnamefont {K.~T.}\
  \bibnamefont {Law}},\ }\href {\doibase 10.1103/PhysRevB.86.184516} {\bibfield
   {journal} {\bibinfo  {journal} {Phys. Rev. B}\ }\textbf {\bibinfo {volume}
  {86}},\ \bibinfo {pages} {184516} (\bibinfo {year} {2012})}\BibitemShut
  {NoStop}%
\bibitem [{\citenamefont {Zhang}\ \emph {et~al.}(2013)\citenamefont {Zhang},
  \citenamefont {Kane},\ and\ \citenamefont {Mele}}]{Zhang2013}%
  \BibitemOpen
  \bibfield  {author} {\bibinfo {author} {\bibfnamefont {F.}~\bibnamefont
  {Zhang}}, \bibinfo {author} {\bibfnamefont {C.~L.}\ \bibnamefont {Kane}}, \
  and\ \bibinfo {author} {\bibfnamefont {E.~J.}\ \bibnamefont {Mele}},\ }\href
  {\doibase 10.1103/PhysRevLett.111.056402} {\bibfield  {journal} {\bibinfo
  {journal} {Phys. Rev. Lett.}\ }\textbf {\bibinfo {volume} {111}},\ \bibinfo
  {pages} {056402} (\bibinfo {year} {2013})}\BibitemShut {NoStop}%
\bibitem [{\citenamefont {Nakosai}\ \emph {et~al.}(2013)\citenamefont
  {Nakosai}, \citenamefont {Budich}, \citenamefont {Tanaka}, \citenamefont
  {Trauzettel},\ and\ \citenamefont {Nagaosa}}]{Nakosai2013}%
  \BibitemOpen
  \bibfield  {author} {\bibinfo {author} {\bibfnamefont {S.}~\bibnamefont
  {Nakosai}}, \bibinfo {author} {\bibfnamefont {J.~C.}\ \bibnamefont {Budich}},
  \bibinfo {author} {\bibfnamefont {Y.}~\bibnamefont {Tanaka}}, \bibinfo
  {author} {\bibfnamefont {B.}~\bibnamefont {Trauzettel}}, \ and\ \bibinfo
  {author} {\bibfnamefont {N.}~\bibnamefont {Nagaosa}},\ }\href {\doibase
  10.1103/PhysRevLett.110.117002} {\bibfield  {journal} {\bibinfo  {journal}
  {Phys. Rev. Lett.}\ }\textbf {\bibinfo {volume} {110}},\ \bibinfo {pages}
  {117002} (\bibinfo {year} {2013})}\BibitemShut {NoStop}%
\bibitem [{\citenamefont {Keselman}\ \emph {et~al.}(2013)\citenamefont
  {Keselman}, \citenamefont {Fu}, \citenamefont {Stern},\ and\ \citenamefont
  {Berg}}]{Keselman2013}%
  \BibitemOpen
  \bibfield  {author} {\bibinfo {author} {\bibfnamefont {A.}~\bibnamefont
  {Keselman}}, \bibinfo {author} {\bibfnamefont {L.}~\bibnamefont {Fu}},
  \bibinfo {author} {\bibfnamefont {A.}~\bibnamefont {Stern}}, \ and\ \bibinfo
  {author} {\bibfnamefont {E.}~\bibnamefont {Berg}},\ }\href {\doibase
  10.1103/PhysRevLett.111.116402} {\bibfield  {journal} {\bibinfo  {journal}
  {Phys. Rev. Lett.}\ }\textbf {\bibinfo {volume} {111}},\ \bibinfo {pages}
  {116402} (\bibinfo {year} {2013})}\BibitemShut {NoStop}%
\bibitem [{\citenamefont {Haim}\ \emph {et~al.}(2014)\citenamefont {Haim},
  \citenamefont {Keselman}, \citenamefont {Berg},\ and\ \citenamefont
  {Oreg}}]{Haim2014}%
  \BibitemOpen
  \bibfield  {author} {\bibinfo {author} {\bibfnamefont {A.}~\bibnamefont
  {Haim}}, \bibinfo {author} {\bibfnamefont {A.}~\bibnamefont {Keselman}},
  \bibinfo {author} {\bibfnamefont {E.}~\bibnamefont {Berg}}, \ and\ \bibinfo
  {author} {\bibfnamefont {Y.}~\bibnamefont {Oreg}},\ }\href {\doibase
  10.1103/PhysRevB.89.220504} {\bibfield  {journal} {\bibinfo  {journal} {Phys.
  Rev. B}\ }\textbf {\bibinfo {volume} {89}},\ \bibinfo {pages} {220504}
  (\bibinfo {year} {2014})}\BibitemShut {NoStop}%
\bibitem [{\citenamefont {Gaidamauskas}\ \emph {et~al.}(2014)\citenamefont
  {Gaidamauskas}, \citenamefont {Paaske},\ and\ \citenamefont
  {Flensberg}}]{Gaidamauskas2014}%
  \BibitemOpen
  \bibfield  {author} {\bibinfo {author} {\bibfnamefont {E.}~\bibnamefont
  {Gaidamauskas}}, \bibinfo {author} {\bibfnamefont {J.}~\bibnamefont
  {Paaske}}, \ and\ \bibinfo {author} {\bibfnamefont {K.}~\bibnamefont
  {Flensberg}},\ }\href {\doibase 10.1103/PhysRevLett.112.126402} {\bibfield
  {journal} {\bibinfo  {journal} {Phys. Rev. Lett.}\ }\textbf {\bibinfo
  {volume} {112}},\ \bibinfo {pages} {126402} (\bibinfo {year}
  {2014})}\BibitemShut {NoStop}%
\bibitem [{\citenamefont {Liu}\ \emph {et~al.}(2014)\citenamefont {Liu},
  \citenamefont {Wong},\ and\ \citenamefont {Law}}]{Liu2014}%
  \BibitemOpen
  \bibfield  {author} {\bibinfo {author} {\bibfnamefont {X.-J.}\ \bibnamefont
  {Liu}}, \bibinfo {author} {\bibfnamefont {C.~L.~M.}\ \bibnamefont {Wong}}, \
  and\ \bibinfo {author} {\bibfnamefont {K.~T.}\ \bibnamefont {Law}},\ }\href
  {\doibase 10.1103/PhysRevX.4.021018} {\bibfield  {journal} {\bibinfo
  {journal} {Phys. Rev. X}\ }\textbf {\bibinfo {volume} {4}},\ \bibinfo {pages}
  {021018} (\bibinfo {year} {2014})}\BibitemShut {NoStop}%
\bibitem [{\citenamefont {W\"olms}\ \emph {et~al.}(2014)\citenamefont
  {W\"olms}, \citenamefont {Stern},\ and\ \citenamefont
  {Flensberg}}]{WolmsStern2014}%
  \BibitemOpen
  \bibfield  {author} {\bibinfo {author} {\bibfnamefont {K.}~\bibnamefont
  {W\"olms}}, \bibinfo {author} {\bibfnamefont {A.}~\bibnamefont {Stern}}, \
  and\ \bibinfo {author} {\bibfnamefont {K.}~\bibnamefont {Flensberg}},\ }\href
  {\doibase 10.1103/PhysRevLett.113.246401} {\bibfield  {journal} {\bibinfo
  {journal} {Phys. Rev. Lett.}\ }\textbf {\bibinfo {volume} {113}},\ \bibinfo
  {pages} {246401} (\bibinfo {year} {2014})}\BibitemShut {NoStop}%
\bibitem [{\citenamefont {Zhao}\ and\ \citenamefont {Wang}(2014)}]{Zhao2014}%
  \BibitemOpen
  \bibfield  {author} {\bibinfo {author} {\bibfnamefont {Y.~X.}\ \bibnamefont
  {Zhao}}\ and\ \bibinfo {author} {\bibfnamefont {Z.~D.}\ \bibnamefont
  {Wang}},\ }\href {\doibase 10.1103/PhysRevB.90.115158} {\bibfield  {journal}
  {\bibinfo  {journal} {Phys. Rev. B}\ }\textbf {\bibinfo {volume} {90}},\
  \bibinfo {pages} {115158} (\bibinfo {year} {2014})}\BibitemShut {NoStop}%
\bibitem [{\citenamefont {Giamarchi}(2004)}]{GiamarchiBook}%
  \BibitemOpen
  \bibfield  {author} {\bibinfo {author} {\bibfnamefont {T.}~\bibnamefont
  {Giamarchi}},\ }\href@noop {} {\emph {\bibinfo {title} {Quantum physics in
  one dimension}}}\ (\bibinfo {year} {2004})\BibitemShut {NoStop}%
\bibitem [{\citenamefont {Altland}\ and\ \citenamefont
  {Zirnbauer}(1997)}]{AltlandZirnbauer1997}%
  \BibitemOpen
  \bibfield  {author} {\bibinfo {author} {\bibfnamefont {A.}~\bibnamefont
  {Altland}}\ and\ \bibinfo {author} {\bibfnamefont {M.~R.}\ \bibnamefont
  {Zirnbauer}},\ }\href {\doibase 10.1103/PhysRevB.55.1142} {\bibfield
  {journal} {\bibinfo  {journal} {Physical Review B}\ }\textbf {\bibinfo
  {volume} {55}},\ \bibinfo {pages} {1142} (\bibinfo {year}
  {1997})}\BibitemShut {NoStop}%
\bibitem [{\citenamefont {Luther}\ and\ \citenamefont
  {Emery}(1974)}]{LutherEmery1974}%
  \BibitemOpen
  \bibfield  {author} {\bibinfo {author} {\bibfnamefont {A.}~\bibnamefont
  {Luther}}\ and\ \bibinfo {author} {\bibfnamefont {V.~J.}\ \bibnamefont
  {Emery}},\ }\href {\doibase 10.1103/PhysRevLett.33.589} {\bibfield  {journal}
  {\bibinfo  {journal} {Phys. Rev. Lett.}\ }\textbf {\bibinfo {volume} {33}},\
  \bibinfo {pages} {589} (\bibinfo {year} {1974})}\BibitemShut {NoStop}%
\bibitem [{com()}]{comment-sg}%
  \BibitemOpen
  \href@noop {} {}\bibinfo {note} {If the higher order cosine terms (higher
  harmonics) are relevant, the transition can become first order, or there
  could be an intermediate phase in which time reversal symmetry is
  spontaneously broken. However, a smooth crossover from the $g<0$ to the $g>0$
  phase is not possible.}\BibitemShut {Stop}%
\bibitem [{\citenamefont {Giamarchi}\ and\ \citenamefont
  {Schulz}(1986)}]{GiamarchiSchulz1986}%
  \BibitemOpen
  \bibfield  {author} {\bibinfo {author} {\bibfnamefont {T.}~\bibnamefont
  {Giamarchi}}\ and\ \bibinfo {author} {\bibfnamefont {H.~J.}\ \bibnamefont
  {Schulz}},\ }\href {\doibase 10.1103/PhysRevB.33.2066} {\bibfield  {journal}
  {\bibinfo  {journal} {Phys. Rev. B}\ }\textbf {\bibinfo {volume} {33}},\
  \bibinfo {pages} {2066} (\bibinfo {year} {1986})}\BibitemShut {NoStop}%
\bibitem [{\citenamefont {{Giamarchi, T.}}\ and\ \citenamefont {{Schulz,
  H.J.}}(1988)}]{GiamarchiSchulz1988}%
  \BibitemOpen
  \bibfield  {author} {\bibinfo {author} {\bibnamefont {{Giamarchi, T.}}}\ and\
  \bibinfo {author} {\bibnamefont {{Schulz, H.J.}}},\ }\href {\doibase
  10.1051/jphys:01988004905081900} {\bibfield  {journal} {\bibinfo  {journal}
  {J. Phys. France}\ }\textbf {\bibinfo {volume} {49}},\ \bibinfo {pages} {819}
  (\bibinfo {year} {1988})}\BibitemShut {NoStop}%
\bibitem [{\citenamefont {{Pollmann}}\ \emph {et~al.}(2010)\citenamefont
  {{Pollmann}}, \citenamefont {{Turner}}, \citenamefont {{Berg}},\ and\
  \citenamefont {{Oshikawa}}}]{Turner2010}%
  \BibitemOpen
  \bibfield  {author} {\bibinfo {author} {\bibfnamefont {F.}~\bibnamefont
  {{Pollmann}}}, \bibinfo {author} {\bibfnamefont {A.~M.}\ \bibnamefont
  {{Turner}}}, \bibinfo {author} {\bibfnamefont {E.}~\bibnamefont {{Berg}}}, \
  and\ \bibinfo {author} {\bibfnamefont {M.}~\bibnamefont {{Oshikawa}}},\
  }\href {\doibase 10.1103/PhysRevB.81.064439} {\bibfield  {journal} {\bibinfo
  {journal} {\prb}\ }\textbf {\bibinfo {volume} {81}},\ \bibinfo {eid} {064439}
  (\bibinfo {year} {2010})},\ \Eprint {http://arxiv.org/abs/0910.1811}
  {arXiv:0910.1811 [cond-mat.str-el]} \BibitemShut {NoStop}%
\bibitem [{\citenamefont {{Turner}}\ \emph {et~al.}(2011)\citenamefont
  {{Turner}}, \citenamefont {{Pollmann}},\ and\ \citenamefont
  {{Berg}}}]{Turner2011}%
  \BibitemOpen
  \bibfield  {author} {\bibinfo {author} {\bibfnamefont {A.~M.}\ \bibnamefont
  {{Turner}}}, \bibinfo {author} {\bibfnamefont {F.}~\bibnamefont
  {{Pollmann}}}, \ and\ \bibinfo {author} {\bibfnamefont {E.}~\bibnamefont
  {{Berg}}},\ }\href {\doibase 10.1103/PhysRevB.83.075102} {\bibfield
  {journal} {\bibinfo  {journal} {\prb}\ }\textbf {\bibinfo {volume} {83}},\
  \bibinfo {eid} {075102} (\bibinfo {year} {2011})},\ \Eprint
  {http://arxiv.org/abs/1008.4346} {arXiv:1008.4346 [cond-mat.str-el]}
  \BibitemShut {NoStop}%
\bibitem [{\citenamefont {Fidkowski}\ and\ \citenamefont
  {Kitaev}(2011)}]{Fidkowski2011a}%
  \BibitemOpen
  \bibfield  {author} {\bibinfo {author} {\bibfnamefont {L.}~\bibnamefont
  {Fidkowski}}\ and\ \bibinfo {author} {\bibfnamefont {A.}~\bibnamefont
  {Kitaev}},\ }\href {\doibase 10.1103/PhysRevB.83.075103} {\bibfield
  {journal} {\bibinfo  {journal} {Phys. Rev. B}\ }\textbf {\bibinfo {volume}
  {83}},\ \bibinfo {pages} {075103} (\bibinfo {year} {2011})}\BibitemShut
  {NoStop}%
\bibitem [{\citenamefont {{Chen}}\ \emph
  {et~al.}(2011{\natexlab{a}})\citenamefont {{Chen}}, \citenamefont {{Gu}},\
  and\ \citenamefont {{Wen}}}]{Chen2011}%
  \BibitemOpen
  \bibfield  {author} {\bibinfo {author} {\bibfnamefont {X.}~\bibnamefont
  {{Chen}}}, \bibinfo {author} {\bibfnamefont {Z.-C.}\ \bibnamefont {{Gu}}}, \
  and\ \bibinfo {author} {\bibfnamefont {X.-G.}\ \bibnamefont {{Wen}}},\ }\href
  {\doibase 10.1103/PhysRevB.83.035107} {\bibfield  {journal} {\bibinfo
  {journal} {\prb}\ }\textbf {\bibinfo {volume} {83}},\ \bibinfo {eid} {035107}
  (\bibinfo {year} {2011}{\natexlab{a}})},\ \Eprint
  {http://arxiv.org/abs/1008.3745} {arXiv:1008.3745 [cond-mat.str-el]}
  \BibitemShut {NoStop}%
\bibitem [{\citenamefont {{Chen}}\ \emph
  {et~al.}(2011{\natexlab{b}})\citenamefont {{Chen}}, \citenamefont {{Gu}},\
  and\ \citenamefont {{Wen}}}]{Chen2011a}%
  \BibitemOpen
  \bibfield  {author} {\bibinfo {author} {\bibfnamefont {X.}~\bibnamefont
  {{Chen}}}, \bibinfo {author} {\bibfnamefont {Z.-C.}\ \bibnamefont {{Gu}}}, \
  and\ \bibinfo {author} {\bibfnamefont {X.-G.}\ \bibnamefont {{Wen}}},\ }\href
  {\doibase 10.1103/PhysRevB.84.235128} {\bibfield  {journal} {\bibinfo
  {journal} {\prb}\ }\textbf {\bibinfo {volume} {84}},\ \bibinfo {eid} {235128}
  (\bibinfo {year} {2011}{\natexlab{b}})},\ \Eprint
  {http://arxiv.org/abs/1103.3323} {arXiv:1103.3323 [cond-mat.str-el]}
  \BibitemShut {NoStop}%
\bibitem [{\citenamefont {{Schuch}}\ \emph {et~al.}(2011)\citenamefont
  {{Schuch}}, \citenamefont {{P{\'e}rez-Garc{\'{\i}}a}},\ and\ \citenamefont
  {{Cirac}}}]{Schuch2011}%
  \BibitemOpen
  \bibfield  {author} {\bibinfo {author} {\bibfnamefont {N.}~\bibnamefont
  {{Schuch}}}, \bibinfo {author} {\bibfnamefont {D.}~\bibnamefont
  {{P{\'e}rez-Garc{\'{\i}}a}}}, \ and\ \bibinfo {author} {\bibfnamefont
  {I.}~\bibnamefont {{Cirac}}},\ }\href {\doibase 10.1103/PhysRevB.84.165139}
  {\bibfield  {journal} {\bibinfo  {journal} {\prb}\ }\textbf {\bibinfo
  {volume} {84}},\ \bibinfo {eid} {165139} (\bibinfo {year} {2011})},\ \Eprint
  {http://arxiv.org/abs/1010.3732} {arXiv:1010.3732 [cond-mat.str-el]}
  \BibitemShut {NoStop}%
\bibitem [{\citenamefont {Chung}\ \emph {et~al.}(2013)\citenamefont {Chung},
  \citenamefont {Horowitz},\ and\ \citenamefont {Qi}}]{Chung2013}%
  \BibitemOpen
  \bibfield  {author} {\bibinfo {author} {\bibfnamefont {S.~B.}\ \bibnamefont
  {Chung}}, \bibinfo {author} {\bibfnamefont {J.}~\bibnamefont {Horowitz}}, \
  and\ \bibinfo {author} {\bibfnamefont {X.-L.}\ \bibnamefont {Qi}},\ }\href
  {\doibase 10.1103/PhysRevB.88.214514} {\bibfield  {journal} {\bibinfo
  {journal} {Phys. Rev. B}\ }\textbf {\bibinfo {volume} {88}},\ \bibinfo
  {pages} {214514} (\bibinfo {year} {2013})}\BibitemShut {NoStop}%
\bibitem [{\citenamefont {White}(1992)}]{White1992PRL}%
  \BibitemOpen
  \bibfield  {author} {\bibinfo {author} {\bibfnamefont {S.~R.}\ \bibnamefont
  {White}},\ }\href {\doibase 10.1103/PhysRevLett.69.2863} {\bibfield
  {journal} {\bibinfo  {journal} {Phys. Rev. Lett.}\ }\textbf {\bibinfo
  {volume} {69}},\ \bibinfo {pages} {2863} (\bibinfo {year}
  {1992})}\BibitemShut {NoStop}%
\bibitem [{\citenamefont {White}(1993)}]{White1992PRB}%
  \BibitemOpen
  \bibfield  {author} {\bibinfo {author} {\bibfnamefont {S.~R.}\ \bibnamefont
  {White}},\ }\href {\doibase 10.1103/PhysRevB.48.10345} {\bibfield  {journal}
  {\bibinfo  {journal} {Phys. Rev. B}\ }\textbf {\bibinfo {volume} {48}},\
  \bibinfo {pages} {10345} (\bibinfo {year} {1993})}\BibitemShut {NoStop}%
\bibitem [{\citenamefont {Schollw{\"o}ck}(2011)}]{SchollwockReview}%
  \BibitemOpen
  \bibfield  {author} {\bibinfo {author} {\bibfnamefont {U.}~\bibnamefont
  {Schollw{\"o}ck}},\ }\href@noop {} {\bibfield  {journal} {\bibinfo  {journal}
  {Annals of Physics}\ }\textbf {\bibinfo {volume} {326}},\ \bibinfo {pages}
  {96} (\bibinfo {year} {2011})}\BibitemShut {NoStop}%
\bibitem [{ITe()}]{ITensor}%
  \BibitemOpen
  \href@noop {} {}\bibinfo {note} {Calculations were performed using the
  ITensor Library, \href{http://itensor.org/}{http://itensor.org/}}\BibitemShut
  {NoStop}%
\bibitem [{\citenamefont {Cardy}(1996)}]{CardyBook}%
  \BibitemOpen
  \bibfield  {author} {\bibinfo {author} {\bibfnamefont {J.}~\bibnamefont
  {Cardy}},\ }\href@noop {} {\emph {\bibinfo {title} {Scaling and
  renormalization in statistical physics}}},\ Vol.~\bibinfo {volume} {5}\
  (\bibinfo  {publisher} {Cambridge University Press},\ \bibinfo {year}
  {1996})\BibitemShut {NoStop}%
\bibitem [{\citenamefont {Qi}\ \emph {et~al.}(2010)\citenamefont {Qi},
  \citenamefont {Hughes},\ and\ \citenamefont {Zhang}}]{Qi2010}%
  \BibitemOpen
  \bibfield  {author} {\bibinfo {author} {\bibfnamefont {X.-L.}\ \bibnamefont
  {Qi}}, \bibinfo {author} {\bibfnamefont {T.~L.}\ \bibnamefont {Hughes}}, \
  and\ \bibinfo {author} {\bibfnamefont {S.-C.}\ \bibnamefont {Zhang}},\ }\href
  {\doibase 10.1103/PhysRevB.81.134508} {\bibfield  {journal} {\bibinfo
  {journal} {Phys. Rev. B}\ }\textbf {\bibinfo {volume} {81}},\ \bibinfo
  {pages} {134508} (\bibinfo {year} {2010})}\BibitemShut {NoStop}%
\bibitem [{\citenamefont {Kestner}\ \emph {et~al.}(2011)\citenamefont
  {Kestner}, \citenamefont {Wang}, \citenamefont {Sau},\ and\ \citenamefont
  {Das~Sarma}}]{Kestner2011}%
  \BibitemOpen
  \bibfield  {author} {\bibinfo {author} {\bibfnamefont {J.~P.}\ \bibnamefont
  {Kestner}}, \bibinfo {author} {\bibfnamefont {B.}~\bibnamefont {Wang}},
  \bibinfo {author} {\bibfnamefont {J.~D.}\ \bibnamefont {Sau}}, \ and\
  \bibinfo {author} {\bibfnamefont {S.}~\bibnamefont {Das~Sarma}},\ }\href
  {\doibase 10.1103/PhysRevB.83.174409} {\bibfield  {journal} {\bibinfo
  {journal} {Phys. Rev. B}\ }\textbf {\bibinfo {volume} {83}},\ \bibinfo
  {pages} {174409} (\bibinfo {year} {2011})}\BibitemShut {NoStop}%
\bibitem [{\citenamefont {{Iemini}}\ \emph {et~al.}(2015)\citenamefont
  {{Iemini}}, \citenamefont {{Mazza}}, \citenamefont {{Rossini}}, \citenamefont
  {{Diehl}},\ and\ \citenamefont {{Fazio}}}]{Iemini2015}%
  \BibitemOpen
  \bibfield  {author} {\bibinfo {author} {\bibfnamefont {F.}~\bibnamefont
  {{Iemini}}}, \bibinfo {author} {\bibfnamefont {L.}~\bibnamefont {{Mazza}}},
  \bibinfo {author} {\bibfnamefont {D.}~\bibnamefont {{Rossini}}}, \bibinfo
  {author} {\bibfnamefont {S.}~\bibnamefont {{Diehl}}}, \ and\ \bibinfo
  {author} {\bibfnamefont {R.}~\bibnamefont {{Fazio}}},\ }\href@noop {}
  {\bibfield  {journal} {\bibinfo  {journal} {ArXiv e-prints}\ } (\bibinfo
  {year} {2015})},\ \Eprint {http://arxiv.org/abs/1504.04230} {arXiv:1504.04230
  [cond-mat.str-el]} \BibitemShut {NoStop}%
\bibitem [{\citenamefont {{Fidkowski}}\ and\ \citenamefont
  {{Kitaev}}(2010)}]{Fidkowski2010}%
  \BibitemOpen
  \bibfield  {author} {\bibinfo {author} {\bibfnamefont {L.}~\bibnamefont
  {{Fidkowski}}}\ and\ \bibinfo {author} {\bibfnamefont {A.}~\bibnamefont
  {{Kitaev}}},\ }\href {\doibase 10.1103/PhysRevB.81.134509} {\bibfield
  {journal} {\bibinfo  {journal} {\prb}\ }\textbf {\bibinfo {volume} {81}},\
  \bibinfo {eid} {134509} (\bibinfo {year} {2010})},\ \Eprint
  {http://arxiv.org/abs/0904.2197} {arXiv:0904.2197 [cond-mat.str-el]}
  \BibitemShut {NoStop}%
\end{thebibliography}%

\end{document}